\documentclass[%
 reprint,
 amsmath,amssymb,
 aps,
 pre,
]{revtex4-2}

\usepackage{graphicx}
\usepackage{dcolumn}
\usepackage{bm}
\usepackage{amssymb}
\usepackage{float}
\usepackage{color}
\usepackage{subfigure}

\usepackage[caption=false]{subfig}

\begin{document}


\title{Local order metrics for many-particle systems across length scales}


\author{Charles Emmett Maher}
 \affiliation{Department of Chemistry, Princeton University, New Jersey 08544, USA}
\author{Salvatore Torquato}%
 \email{torquato@princeton.edu}
\affiliation{Department of Chemistry, Princeton University, Princeton, New Jersey 08544, USA}
\affiliation{Department of Physics, Princeton University, Princeton, New Jersey 08544, USA}
\affiliation{Princeton Institute for the Science and Technology of Materials, Princeton University, Princeton, New Jersey 08544, USA}
\affiliation{Program in Applied and Computational Mathematics, Princeton University, Princeton, New Jersey 08544, USA}

\date{\today}

\begin{abstract}
Formulating order metrics that sensitively quantify the degree of order/disorder in many-particle systems in $d$-dimensional Euclidean space $\mathbb{R}^d$ across length scales is an outstanding challenge in physics, chemistry, and materials science. 
Since an infinite set of $n$-particle correlation functions is required to fully characterize a system, one must settle for a reduced set of structural information, in practice. 
We initiate a program to use the local number variance $\sigma_N^2(R)$ associated with a spherical sampling window of radius $R$ (which encodes pair correlations) and an integral measure derived from it $\Sigma_N(R_i,R_j)$ that depends on two specified radial distances $R_i$ and $R_j$. 
Across the first three space dimensions ($d = 1,2,3$), we find these metrics can sensitively describe and categorize the degree of order/disorder of 41 different models of antihyperuniform, nonhyperuniform, disordered hyperuniform, and ordered hyperuniform many-particle systems at a specified length scale $R$.
Using our local variance metrics, we demonstrate the importance of assessing order/disorder with respect to a specific value of $R$.
These local order metrics could also aid in the inverse design of structures with prescribed length-scale-specific degrees of order/disorder that yield desired physical properties.
In future work, it would be fruitful to explore the use of higher-order moments of the number of points within a spherical window of radius $R$ [S. Torquato {\it et al}., Phys. Rev. X, \textbf{11}, 021028 (2021)] to devise even more sensitive order metrics.
\end{abstract}

\maketitle

\section{Introduction} 
The classification of both ordered and disordered many-particle systems is a difficult and outstanding problem in physics, chemistry, and materials science.
To fully characterize the microstructure of a many-particle system in $d$-dimensional Euclidean space $\mathbb{R}^d$, one requires an infinite set of $n$-particle probability density functions $\rho_n(\mathbf{r}^n)$ associated with finding \textit{any} $n$ particles in a configuration $\mathbf{r}^n$ \cite{Ha86, To02}.
Such complete information is virtually never available in practice, so we \textit{must settle for reduced structural information}, e.g., by only considering lower-order correlation functions.
Subject to this \textit{necessary} restriction, several scalar order metrics (see, e.g., Refs. \cite{To00_2, Tr00, Ka02, To10, To18_2, Zh15}) have been proposed that have been fruitfully applied to arrangements of nonoverlapping spheres (sphere packings) \cite{Tr00, To00_2, To03, Ka02, Tm19}, simple liquids \cite{Ri96, Tr00, Ri96_2, Er03}, glasses \cite{Tr00, Zh16}, water \cite{Er01, Er02, Ga21}, disordered ground states \cite{Zh15, Mo23}, random media \cite{Di18, Kl19, Ba19, Kl20, Wa24}, the prime numbers \cite{To19}, two-phase media \cite{To22}, and random fields \cite{Sk24}.


Previous work has advocated the need for order metrics that apply across length scales\cite{Ka02, To10, To18_2, To22, Sk24}.
Motivated by the success of analogous variance-based order metrics for two-phase media and random scalar fields \cite{To22, Sk24, F1}, and following a suggestion from Ref. \citenum{To03}, we aim to characterize the degree of order/disorder in a general many-particle system in $\mathbb{R}^d$, at any length scale, using the local number variance $\sigma_N^2(R)$ associated with a spherical window of radius $R$ (which encodes pair correlations).
The local number variance quantifies local number density fluctuations, which are of fundamental importance to a plethora of physical, mathematical, and biological disciplines \cite{Sc66, Ve75, Zi77, Ca78, Ha86, Jo91, Pe93, Bl93, Tr98, To00, Ga02, Wa02, To03, La03, Ma04, Kl16, To18, Kl21, Ph23, Ma23}.

The local number variance can be obtained via direct sampling of point configurations, i.e., $\sigma_N^2(R)\equiv\langle N^2(R)\rangle - \langle N(R)\rangle^2$, where $N(R)$ is the number of points in a spherical window of radius $R$, or given in terms of the pair correlation function $g_2(\mathbf{r})$ in direct space or the structure factor $S(\mathbf{k})$ in reciprocal space (see Sec. II) \cite{To03}.
The large-$R$ behavior of $\sigma_N^2(R)$ is central to the hyperuniformity concept \cite{To03, To18}.
In particular, a hyperuniform many-particle system is one in which $\sigma_N^2(R)$ grows more slowly than the window volume $R^d$, i.e. \cite{To03, To18},
\begin{equation}\label{eq:hu_nv}
    \lim_{R\rightarrow\infty}\frac{\sigma_N^2(R)}{v_1(R)}=0,
\end{equation}
where 
\begin{equation}\label{eq:v1}
    v_1(R) = \frac{\pi^{d/2}R^d}{\Gamma(d/2 + 1)},
\end{equation}
is the volume of a $d$-dimensional sphere of radius $R$ and $\Gamma(x)$ is the gamma function.
In other words, the system is characterized by large-scale density fluctuations that are anomalously suppressed compared to those of a typical disordered system.
Equivalently, a system is hyperuniform if $S(\mathbf{k})\rightarrow 0$ as $|\mathbf{k}|\rightarrow0$.

The hyperuniformity concept generalizes the traditional notion of long-range order of crystals and quasicrystals to also contain exotic disordered states of matter \cite{To03, To18}.
Moreover, it offers a unified means to classify equilibrium and nonequilibrium many-particle systems, whether hyperuniform or not, according to their large-scale fluctuation characteristics.
Suppose the structure factor has the following power-law form as $|\mathbf{k}|$ tends to zero:
\begin{equation}\label{eq:alphadef}
    S(\mathbf{k})\sim|\mathbf{k}|^{\alpha}\quad(|\mathbf{k}|\rightarrow0)
\end{equation}
where $\alpha$ is the \textit{hyperuniformity scaling exponent}.
Hyperuniform many-particle systems, which have $\alpha > 0$, can be divided into three distinct classes based on $\alpha$ that describe their associated large-$R$ behaviors of $\sigma_N^2(R)$ \cite{To03, To18}.
Specifically, in order of decreasing ``strength,'' there are class I ($\alpha > 1$), class II ($\alpha = 1$), and class III ($\textcolor{black}{0} < \alpha < 1$) hyperuniform systems.
By contrast, ``typical'' nonhyperuniform many-particle systems have bounded, positive $S(0)$ ($\alpha = 0$) \cite{To18}, and \textit{anti}hyperuniform systems have unbounded $S(0)$ ($-d < \alpha < 0$) \cite{To21}.

In this work, we initiate a program to use the scaled local number variance $\sigma_N^2(R)/v_1(R)$ and an integral measure derived from it $\Sigma_N(R_i,R_j)$ that depends on two specified radial distances $R_i$ and $R_j$ as an order metric for antihyperuniform, nonhyperuniform, disordered hyperuniform, and ordered hyperuniform point configurations at any length scale $R$.
Specifically, at a specified $R$, lower (higher) number density fluctuations as measured by $\sigma_N^2(R)$ are used as a measure of a greater degree of order (disorder).
Following Refs. \citenum{To22} and \citenum{Sk24}, which profitably use a $1/v_1(R)$ scaling in the formulation of their $\sigma_V^2(R)$- and $\sigma_F^2(R)$-based order metrics, we demonstrate that $1/v_1(R)$ is also a good choice of scale for a $\sigma_N^2(R)$-based metric.
Across the first three space dimensions ($d=1,2,$ and 3) we find that $\sigma_N^2(R)/v_1(R)$ and $\Sigma_N(R_i,R_j)$ can sensitively describe the degree of order/disorder of 41 different models of the aforementioned broad spectrum of many-particle systems at short, intermediate, and large length scales consistently with physical intuition.
While many existing translational order metrics (see, e.g., Refs. \citenum{To10,To18_2} and references therein) attempt to quantify the degree of order/disorder of a system globally (i.e., not at some selected length scale), our local variance metrics depend explicitly on a length scale $R$, and thus can sensitively and robustly characterize order/disorder on any prescribed length scale.
We compare and contrast our number variance-based metrics and existing ones in Sec. V.
Moreover, we demonstrate the length scale dependence of order/disorder rankings across models in a given space dimension, providing additional evidence that it is important to assess order/disorder with respect to a specific length scale \cite{To22,Sk24}.
To aid in materials design, these order metrics could be applied to inverse techniques \cite{Re07, To09, Di13, Tr20} to generate structures with prescribed scale-specific degrees of order/disorder, yielding desired physical properties.

The rest of the paper is structured as follows. Section II contains pertinent background information and mathematical definitions. Section III describes the large variety of systems to which we apply our order metrics. In Sec. IV we present our results, and in Sec. V we offer our conclusions and potential areas for future study.

\section{Background and Definitions}
\subsection{Pair statistics of many-particle systems}
Consider a statistically homogeneous (translationally invariant) point process in $\mathbb{R}^d$ with a number density $\rho$. 
It is completely statistically characterized by specifying the the countably infinite set of $n$-particle probability density functions $\rho_n(\mathbf{r}^n)$, where $\mathbf{R}^N \equiv {\mathbf{r}^1,\mathbf{r}^2,\dots,\mathbf{r}^N}$ \cite{To06}.
It is convenient to define the so-called $n$-particle correlation function,
\begin{equation}
    g_n(\mathbf{r}^N)=\frac{\rho_n(\mathbf{r}^n)}{\rho^N}.
\end{equation}

The important two-particle quantity
\begin{equation}\label{eq:pcf}
    g_2(\mathbf{r}_{12})=\frac{\rho_2(\mathbf{r}_{12})}{\rho^2}
\end{equation}
where $\mathbf{r}_{ij}=\mathbf{r}_i-\mathbf{r}_j$, is the \textit{pair correlation function}.
Henceforth, we drop the subscript $12$ and simply denote the displacement vector between two bodies as $\mathbf{r}$.
The total correlation function $h(\mathbf{r}_{12})$ is defined as
\begin{equation}
    h(\mathbf{r}) = g_2(\mathbf{r})-1,
\end{equation}
which is zero when there are no spatial correlations.
When the system is isotropic, $g_2(\mathbf{r})$ depends only on the radial distance $r$ between two bodies, i.e.,
\begin{equation}
    g_2(\mathbf{r}) = g_2(r),
\end{equation}
and is called the radial distribution function.

The structure factor $S(k)$ is related to the Fourier transform of $h(\mathbf{r})$ [$\Tilde{h}(\mathbf{r})$] via
\begin{equation}
    S(\mathbf{k}) = 1+\rho\Tilde{h}(\mathbf{r}),
\end{equation}
where $\mathbf{k}$ is the wave vector.
It is well-known that the structure factor is proportional to the scattered intensity of radiation from a many-particle system and be obtained in practice with scattering experiments. 
For statistically isotropic systems, $S(\mathbf{k})$ only depends on the wavenumber $|\mathbf{k}|\equiv k$.

\begin{figure*}[!t]
    \centering
    \subfigure[]{\includegraphics[width=0.3\textwidth]{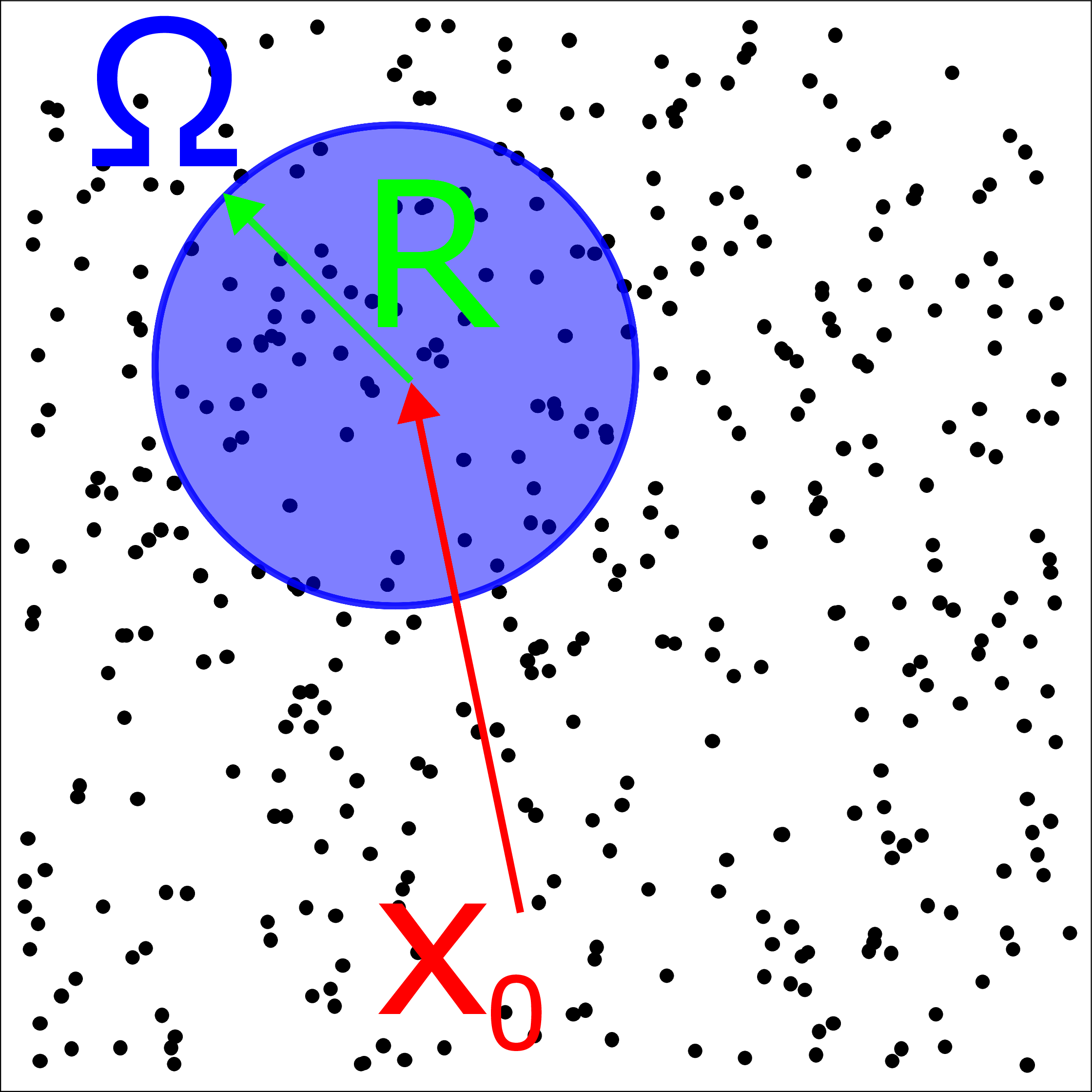} }
    \subfigure[]{\includegraphics[width=0.3\textwidth]{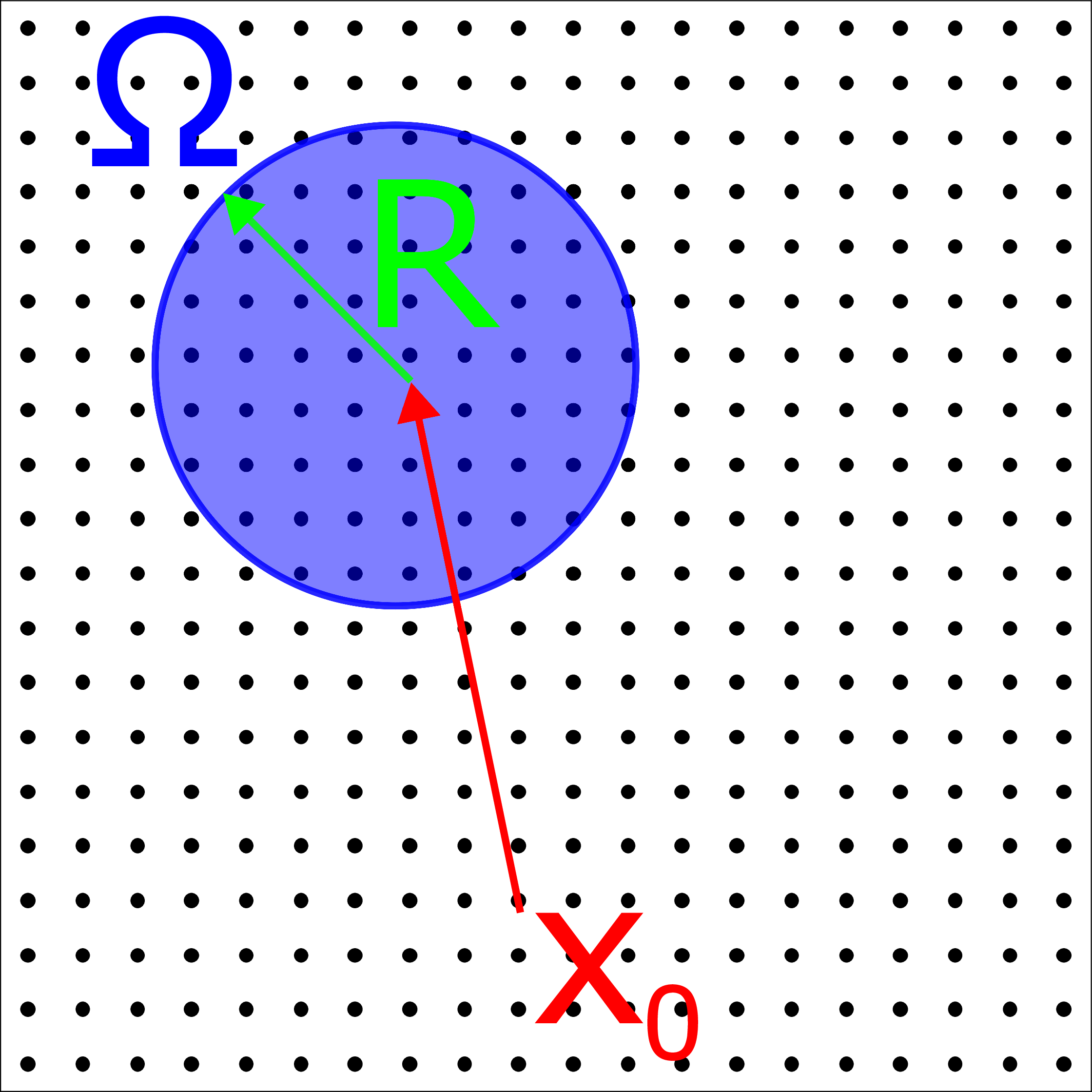} }
    \subfigure[]{\includegraphics[width=0.3\textwidth]{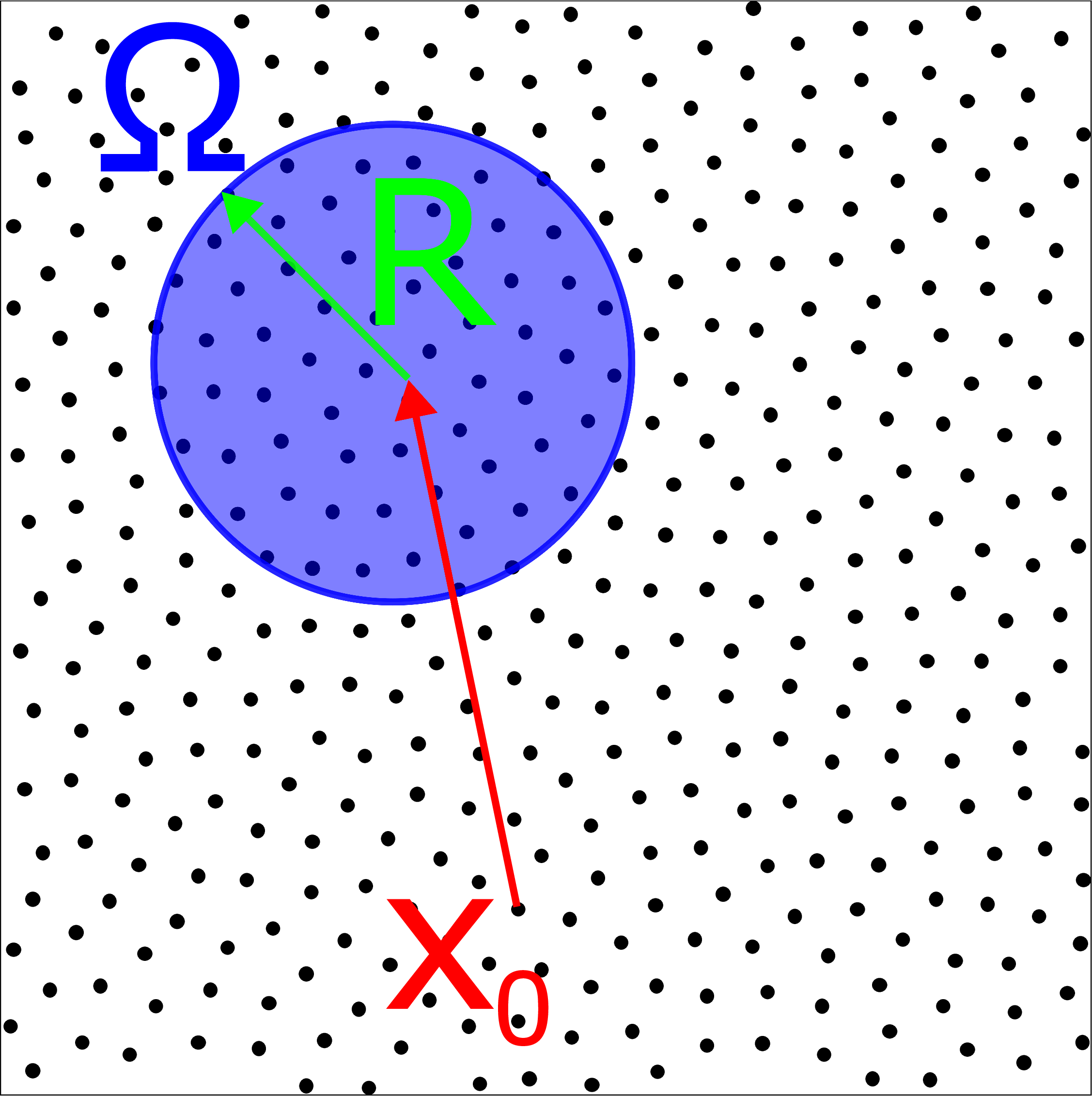} }
    \caption{Schematics indicating a circular observation window $\Omega$ of radius $R$ in two dimensions and its centroid $x_0$ for disordered nonhyperuniform (left), ordered hyperuniform (center), and disordered hyperuniform (right) point patterns. In each example, the number of points within the window $\Omega$ will fluctuate as the window position $x_0$ varies. The number variance $\sigma_N^2(R)$ will asymptotically scale like $R^2$ for the nonhyperuniform point pattern (right) and like $R$ for the ordered (center) and disordered (right) hyperuniform patterns; see Sec. II B for the other possible large-$R$ scalings.}
    \label{fig:schematic}
\end{figure*}

\subsection{Number variance and hyperuniformity of many-particle systems}\label{sec:HU}
Given a point configuration in $\mathbb{R}^d$, consider sampling the number of points $N(R)$ within a $d$-dimensional spherical window of radius $R$. 
The local number variance can be obtained via direct sampling of a many-particle system, i.e., $\sigma_N^2(R)\equiv\langle N^2(R)\rangle - \langle N(R)\rangle^2$, or given in terms of the pair correlation function $g_2(\mathbf{r})$ in direct space or the structure factor $S(\mathbf{k})$ in reciprocal space \cite{To03}:
\begin{equation}\label{eq:nv}
    \begin{split}
        \sigma_N^2 &=\rho v_1(R)\left[1+\rho\int_{\mathbb{R}^d} h(\mathbf{r})\alpha_2(r;R)d\mathbf{r}\right], \\
        &=\rho v_1(R)\left[\frac{1}{(2\pi)^d}\int_{\mathbb{R}^d}S(\mathbf{k})\Tilde{\alpha}_2(k;R)d\mathbf{k} \right],
    \end{split}
\end{equation}
where 
\begin{equation}\label{a2}
    \alpha_2(r;R) =
    \begin{cases}
        1 - \frac{r}{2R},&\;d=1\\
        \frac{2}{\pi}\left[\textrm{cos}^{-1}\left(\frac{r}{2R}\right)-\frac{r}{2R}\left(1-\frac{r^2}{4R^2}\right)^{1/2}\right],&\;d=2\\
        1-\frac{3}{4}\frac{r}{R}+\frac{1}{16}\left(\frac{r}{R}\right)^3,&\;d=3
    \end{cases}
\end{equation}
is the intersection volume of two $d$-dimensional spherical windows of radius R, scaled by the volume of a $d$-dimensional sphere $v_1(R)$ [given by Eq. (\ref{eq:v1})] whose centers are separated by a distance $r$ \cite{To03, To06}, and $\Tilde{\alpha}_2(k;R)$ is its Fourier transform.
Figure \ref{fig:schematic} depicts schematically how one can compute the number variance $\sigma_N^2(R)$ directly from general point patterns by window sampling at some fixed radius $R$.
We describe the computation of $\sigma_N^2(R)$ via direct sampling of a point pattern in more detail in the Appendix.

For asymptotically large $R$, Eq. (\ref{eq:nv}) can be written as \cite{To03, To18}
\begin{equation}\label{eq:nv_asy}
\begin{split}
    \sigma_N^2(R)=2^d\eta \Bigg{[}&A\left(\frac{R}{D}\right)^d + B\left(\frac{R}{D}\right)^{d-1}\\
    + &o\left(\frac{R}{D}\right)^{d-1}\Bigg{]}\;(R\rightarrow+\infty),
\end{split}
\end{equation}
where $\eta = \rho v_1(D/2)$ the dimensionless density, $D$ is a characteristic microscopic length scale, and $o(x)$ signifies all terms of less than order $x$.
$A$ and $B$ are $d$-dependent asymptotic coefficients that multiply terms proportional to the window volume ($R^d$) and window surface area ($R^{d-1}$), respectively.
These coefficients are given explicitly by \cite{To03, To18}
\begin{equation}\label{eq:A}
    A = \lim_{|\mathbf{k}|\rightarrow0}S(\mathbf{k}) = 1 + \rho\int_{\mathbb{R}^d}h(\mathbf{r})d\mathbf{r},
\end{equation}
and 
\begin{equation}\label{eq:B}
    B = -\frac{d\Gamma(d/2)\rho}{2\pi^{1/2}D\Gamma[(d+1)/2]}\int_{\mathbb{R}^d}|\mathbf{r}|h(\mathbf{r})d\mathbf{r}.    
\end{equation}
A Fourier representation for $B$ also exists, which is convenient for computational purposes when $S(k)$ is known, given by \cite{To21}
\begin{equation}
    B=\int_0^{\infty}\frac{S(k)-S(0)}{k^2}dk.
\end{equation}
A many-particle system is hyperuniform if $\sigma_N^2(R)$ grows more slowly than the window volume $R^d$ [see Eq. (\ref{eq:hu_nv})], or, equivalently, if $A = \lim_{|\mathbf{k}|\rightarrow0}S(\mathbf{k}) = 0$ \cite{To03, To18}.

The hyperuniformity exponent $\alpha$ [see Eq. (\ref{eq:alphadef})], can be used to divide hyperuniform systems into three distinct classes that describe their associated large-$R$ behaviors of $\sigma_N^2(R)$ \cite{To03, To18}:
\begin{equation}
    \sigma_N^2(R)\sim
    \begin{cases}
        R^{d-1},\quad\alpha > 1 &\textrm{(Class I)}\\
        R^{d-1}\textrm{ln}(R),\quad\alpha = 1&\textrm{(Class II)}\\
        R^{d-\alpha},\quad0<\alpha<1&\textrm{(Class III)},
    \end{cases}
\end{equation}
where class I and class III are the strongest and weakest forms of hyperuniformity, respectively.
By contrast, nonhyperuniform many-particle systems have the following large-$R$ scaling behaviors \cite{To21}:
\begin{equation}
        \sigma_N^2(R)\sim
    \begin{cases}
        R^{d},\quad\alpha = 0 \;\textrm{(typical nonhyperuniform)}\\
        R^{d-\alpha},\quad-d<\alpha<0\;\textrm{(antihyperuniform)}.
    \end{cases}
\end{equation}
For a typical nonhyperuniform system, $S(0)$ is bounded \cite{To18}, and for \textit{anti}hyperuniform systems, $S(0)$ is unbounded, i.e.,
\begin{equation}
    \lim_{|\mathbf{k}|\rightarrow0}S(\mathbf{k})=+\infty.
\end{equation}
In summary, the scalings (15) and (16) provide a complete characterization of large-scale density fluctuations that span all configurations, hyperuniform or not or ordered or not \cite{To21}.

\subsection{Translational order metrics}
Naturally, there is an enormous family of scalar functions that depend on the particle positions that could in principle be used as an order metric, but they are not necessarily \textit{good} order metrics.
Based on experiences with existing order metrics, Kansal {\it et al}. \cite{Ka02} suggested that a good order metric is one that (1) is sensitive to any type of ordering in a system and should not be biased toward any reference system, (2) reflects the hierarchy of ordering between prototypical systems given by physical intuition, (3) detects order at any length scale, and (4) is affected by the variety of local configurational patterns and the spatial distribution of these patterns.

Both translational and orientational order metrics have been considered in previous works (see, e.g., Refs. \citenum{To10,To18_2} and references therein), but we restrict our discussion here only to translational order metrics, which have been profitably used to characterize many-particle systems.

The order metric $T$ measures the degree of translational order in a system of interest relative to the perfect face-centered-cubic (fcc) lattice structure at the same number density \cite{To00, Tr00}.
Specifically, 
\begin{equation}
    T=\left|\frac{\sum_{i=1}^{N_C}(n_i-n_i^{ideal})}{\sum_{i=1}^{N_C}(n_i^{FCC}-n_i^{ideal})}\right|,
\end{equation}
where $n_i$ (for the system of interest) indicates the average occupation number for the thin spherical shell located a distance away from a reference particle equal to the $i$th nearest-neighbor separation for the fcc lattice at that number density and $N_C$ is the total number of shells considered.
Similarly, $n_i^{ideal}$ and $n_i^{fcc}$ are the corresponding shell occupation numbers for the Poisson point process and the fcc lattice, respectively. 
$T$ notably does not meet the aforementioned property (1) of a good order metric because it explicitly requires comparison to the fcc system.

The $T^*$ order metric, by contrast, is defined independent of any reference lattice structure and provides a measure of the local-density modulations in collections of nonoverlapping spheres of diameter $\mathcal{D}$ \cite{Tr00}
\begin{equation}
    T^* = \frac{\int_{\rho^{1/d}\mathcal{D}}^{\xi_C}|h(\xi)|d\xi}{\xi_C-\rho^{1/d}\mathcal{D}},
\end{equation}
where $\xi=r\rho^{1/d}$ and $\xi_C$ is some cutoff distance.
Density-density correlations are detected by integrating over the absolute value of the total correlation function.
We note also that $T^*$ was designed to assess the degree of order/disorder globally for sphere packings specifically, and thus, its application is limited in scope.

Truskett {\it et al}. \cite{Tr00} also proposed using the excess two-particle entropy $s^{(2)}$ \cite{Ge52, Ne58, Ba89} as an order metric, given by
\begin{equation}\label{eq:s2}
    s^{(2)} = -\frac{k_B\rho}{2}\int_{\mathbb{R}^d}d\mathbf{r}\{g(r)\textrm{ln}(g(r))-[g(r)-1]\},
\end{equation}
where $k_B$ is Boltzmann's constant.
This metric is essentially the multiparticle correlation function expansion of the excess entropy---relative to an ideal gas at the same number density---truncated at pair statistics.
Specifically, the positive semidefinite quantity $-s^{(2)}/k$ is used as an order metric.

Torquato {\it et al}. \cite{Zh15} defined another order metric involving an integral over $h(r)$, specifically, 
\begin{equation}\label{eq:tau}
    \tau=\frac{1}{\rho^{1/d}}\int_{\mathbb{R}^d}h^2(r)d\mathbf{r}=\frac{1}{(2\pi)^d\rho^{1/d}}\int_{\mathbb{R}^d}\Tilde{h}^2(k)d\mathbf{k}.
\end{equation}
Similarly to $T^*$, density-density correlations are detected by integrating over the total correlation function, but with a different weighting.
Such a metric is advantageous over those mentioned previously because it can be formulated in terms of direct space or reciprocal space pair statistics.
In addition, a local version of $\tau$ in reciprocal space, denoted by $\tau(K)$, has been formulated by introducing an upper bound on the integration \cite{Kl19}:
\begin{equation}\label{eq:tauk}
    \tau(K)=\frac{1}{(2\pi)^d\rho^{1/d}}\int_{0}^K\Tilde{h}^2(k)d\mathbf{k}.
\end{equation}
We note that the similarity in construction and ability to quantify order/disorder between $\tau, s^{(2)},$ and $T^*$ has been pointed out in several previous works \cite{Ph23, Zh15, Lo17}.

\section{Hyperuniform and nonhyperuniform Models}\label{sec:models}
We consider 12 different models of statistically homogeneous (translationally invariant) many-particle systems in one, two, and three dimensions: eight nonhyperuniform models, two of which are antihyperuniform, and four hyperuniform models.
Our nonhyperuniform models are chosen such that they span a representative subset of the possible range of $S(0)$ values and our hyperuniform models are chosen such that they span disordered, ordered, stealthy (defined below), and nonstealthy varieties. 
A set of figures depicting representative configurations and pair statistics for each model for $d = 1,2,3$ is given in the Supplemental Material \cite{SI}.
To compare each model in a given space dimension consistently, we scale all distances by $\rho^{1/d}$, which, unless stated otherwise, is a reasonable estimate of the mean nearest-neighbor distance in the system.

\subsection{Hyperuniform models}

\subsubsection{Periodic systems}
Consider a Bravais lattice $\mathcal{L}$ in $\mathbb{R}^d$ in which a single particle is placed in a fundamental cell $\mathcal{F}$ of $\mathcal{L}$.
The structure factor of such a system is given by \cite{To18}
\begin{equation}\label{eq:sklattice}
    S(\mathbf{k})=\frac{(2\pi)^d}{V_{\mathcal{F}}}\sum_{q\in\mathcal{L}^*\backslash {0}}\delta(\mathbf{k}-\mathbf{q})
\end{equation}
where $\mathcal{L}^*$ denotes the reciprocal lattice of $\mathcal{L}$ and $\delta(x)$ is the Dirac delta function.
The structure factor of all Bravais lattices for all wavenumbers up to the first Bragg peak is identically 0, which is referred to as \textit{stealthy} hyperuniformity and makes them class I hyperuniform with $\sigma_N^2(R)\sim R^{d-1}$ scaling in the large-$R$ limit \cite{Uc04, Ba08, Zh15, To18}.
Specifically, we examine the one, two and three-dimensional (1D, 2D, and 3D) hypercubic lattices $\mathbb{Z}^d$, the 2D and 3D root lattices $A_d$ [$A_3$ is also known as the face-centered-cubic (fcc) lattice] and the dual lattice to the fcc lattice $A_3^*$ [or the body-centered-cubic (bcc) lattice].
Here, we use Eqs. (\ref{eq:sklattice}) and (\ref{eq:nv}) to compute the number variance for the set of lattices listed above.
While periodic systems are not statistically homogeneous, the exact calculations of their number variances only involve their radial statistics \cite{Ha86, To18}.

\subsubsection{Quasiperiodic systems}
We also consider 1D and 2D quasiperiodic many-particle systems that have long-ranged orientational order, but exhibit quasiperiodic rather than periodic translational order \cite{St87}.
The structure factors of such systems are composed of a dense set of Bragg peaks separated by gaps of arbitrarily small size \cite{Le84}.
For $d=1$, we examine the Fibonacci quasiperiodic chain, which has $\alpha = 3$ \cite{Og17} and for $d=2$, we examine the quasiperiodic Penrose tiling, which has $\alpha = 6$ \cite{Fu19, Hi24}.
Both these models are class I hyperuniform and thus have $\sigma_N^2(R)\sim R^{d-1}$ scaling in the large-$R$ limit.
Here, for the 1D Fibonacci and 2D Penrose quasicperiodic systems, we directly sample the number variance using configurations generated via the substitution tiling method with $N = 832040$ particles described in Ref. \cite{Og19} and generated from the generalized dual method with $N = 167761$ described in Refs. \citenum{De81, So85, Li17}, respectively.

\subsubsection{Disordered stealthy hyperuniform systems}
Disordered stealthy hyperuniform many-particle systems are class I hyperuniform with $\sigma_N^2(R)\sim R^{d-1}$ scaling in the large-$R$ limit and have structure factors that vanish in a spherical region around the origin, i.e., $S(\mathbf{k}) = 0$ for $0 < \mathbf{k} \leq K$, but, unlike periodic systems, are isotropic and \textit{do not} have Bragg peaks.
Such systems can be generated using the collective-coordinate optimization scheme that involves finding the highly degenerate ground states of a class of bounded pair potentials with compact support in Fourier space, which are stealthy and hyperuniform by construction \cite{Fa91, Uc04, Uc06, Ba08, Zh15, Zh15_2, To18, Mo23}.
The parameter $\chi$ is a dimensionless measure of the ratio of constrained degrees of freedom (i.e., the number of wave vectors contained within the cutoff wavenumber $K$) to the total degrees of freedom (roughly $dN$, where $N$ is the total number of points in the system).
Many-particle systems with small $\chi$ (relatively unconstrained) will have short-range disorder, and as $\chi$ increases, the degree of short-range order increases within the ``disordered regime'' ($\chi < 1/2$ for $d = 2$ and 3 \cite{Zh15}).
For $d=1$, the disordered regime only extends to $\chi < 1/3$ \cite{Zh15_2}.
Here, we directly sample the number variance for ``entropically favored'' disordered stealthy hyperuniform many-particle systems generated via the procedure in Ref. \citenum{Zh15_2}.
In particular, we examine ``high-$\chi$'' stealthy hyperuniform systems, i.e., $\chi = 0.3$ for $d=1$ and $\chi = 0.49$ for $d=2$ and 3.
For $d=1$, we consider 900 configurations with $N = 1000$, for $d=2$, we consider 700 configurations with $N = 10^4$, and for $d=3$, we consider 1000 configurations with $N = 8000$.

\subsubsection{Uniformly randomized lattices}
It has been demonstrated \cite{Kl20} that one can ``cloak'' the long-ranged order in a lattice---meaning the long-range order cannot be reconstructed from the pair-correlation function alone---by applying certain independent and identically distributed perturbations to each lattice site.
Such a protocol yields many-particle systems called cloaked uniformly randomized lattices (URL).
Here, we consider a cloaked URL derived from the hypercubic lattices $\mathbb{Z}^d$, which can be generated by displacing each lattice point by a random vector uniformly distributed on a scaled fundamental cell of the lattice, i.e., $b\mathcal{F}\equiv [-b/2,b/2)^d$, where $b$ controls the perturbation strength and results in cloaking if it is an integer.
Specifically, we choose $b = 1$.
This type of perturbation results in a class I hyperuniform system with $\alpha = 2$ and $\sigma_N^2(R)\sim R^{d-1}$ scaling in the large-$R$ limit.
The structure factor of this URL is given by \cite{Wa22}
\begin{equation}\label{eq:URL}
    S(\mathbf{k}) = 1-2^d\prod_{i-1}^d\frac{\textrm{sin}(k_i/2)}{k_i}.
\end{equation}
We use Eqs. (\ref{eq:URL}) and (\ref{eq:nv}) to compute the number variance for the URL described above for $d=1,2$ and 3.

\subsection{Nonhyperuniform models}
\subsubsection{Poisson point process}
A homogeneous Poisson point process in $\mathbb{R}^d$ is a many-particle system in which the position of each particle is totally independent. 
At unit mean number density ($\rho = 1$), this process can be generated in a fixed hypercubic simulation box with volume $V$ by choosing a random number $N$ from a Poisson distribution with a mean of $\rho V = V$ and then uniformly placing $N$ points in the simulation box.
Such a system with no spatial correlations has 
\begin{equation}\label{eq:Poi}
    g_2(r) = 1\;\textrm{and}\;S(k) = 1,
\end{equation}
and thus, is nonhyperuniform with $\sigma_N^2(R)\sim R^{d}$ scaling in the large-$R$ limit.
We use Eqs. (\ref{eq:Poi}) and (\ref{eq:nv}) to compute the local number variance for $d = 1,2$ and 3.

\subsubsection{Equilibrium Packings}
We consider equilibrium (Gibbs) ensembles of identical nonoverlapping spheres of radius $a$ at a packing fraction $\phi$ across the first three space dimensions \cite{Ha86, To02}.
Specifically, we consider disordered packings along the stable disordered fluid branch in the phase diagram, which are all nonhyperuniform with $\sigma_N^2(R)\sim R^{d}$ scaling in the large-$R$ limit \cite{To02, To18_2}.
The pair statistics for equilibrium 1D hard rods packings are known exactly \cite{Pe64}.
In particular, using the exact solution of the direct correlation function \cite{Pe64, Ze27} and the Ornstein-Zernike integral equation, we can express the structure factor as
\begin{equation}\label{eq:1Deq}
    S(k) = \left[1-\frac{2\phi\{\phi[\textrm{cos}(2ak)-1]+2ak\;\textrm{sin}(2ak)(\phi-1)\}}{(1-\phi)^2(2ak)^2}\right]^{-1}.
\end{equation}
For 3D sphere packings, we use the Percus-Yevick approximation of the structure factor $S(k)$ \cite{Ha86}:
\begin{equation}\label{eq:3Deq}
    \begin{split}
        S(k) = &\bigg{(}1-\rho\frac{16\pi a^3}{q^6}\{[24a_1\phi-12(a_1+2a_2)\phi q^2+\\
        &(12a_2\phi +2a_1+a_1\phi)q^4]\textrm{cos}(q)+[24a_1\phi q-\\
        &2(a_1+2a_1\phi+12a_2\phi)q^3]\textrm{sin}(q)-24\phi(a_1-a_2q^2)\}\bigg{)}^{-1}.
    \end{split}
\end{equation}
where $q=2ka$, $a_1=(1+2\phi)^2/(a-\phi)^4$, and $a_2=-(1+\phi/2)^2/(1-\phi)^4$.
We substitute Eq. (\ref{eq:1Deq}) with $\phi = 0.85$ for $d=1$ and Eq. (\ref{eq:3Deq}) with $\phi = 0.48$ for $d=3$ into Eq. (\ref{eq:nv}) to compute their respective number variances.
There is no closed-form approximation for $S(k)$ for $d=2$, so we directly sample the number variance from 1000 disk packings with $\phi = 0.65$ and $N = 10^4$ generated by a Metropolis numerical scheme \cite{Ha86, To02}.

\subsubsection{Random sequential addition packings}
The random sequential addition (RSA) process is a time-dependent, nonequilibrium procedure that generates disordered sphere packings in $\mathbb{R}^d$ \cite{To06_3, Zh13, To06_2, Re58, Wi66, Fe80, Co88}.
Starting from an initially empty box in $\mathbb{R}^d$, the RSA process is produced by randomly, irreversibly, and sequentially placing nonoverlapping spheres into the box. 
This procedure is repeated for successively larger volumes until an appropriate infinite-volume limit is obtained.
In practice, hard spheres are randomly and sequentially placed in a large periodic simulation box, subject to the nonoverlap constraint, i.e., a sphere is only added to the box if it would not overlap any existing sphere, otherwise it is discarded.
In principle, the RSA process can be stopped at any time $t$, at which point the packing fraction is $\phi(t)$.
Here, we are interested in the infinite-time, maximally saturated limit where $\phi_s = \phi(\infty)$, meaning no additional sphere can be added to the system without causing overlaps \cite{To03}.
Previous numerical calculations indicate that $S(0)$ for saturated RSA packings for $d \leq 7$ is small, but nonzero, meaning they are nonhyperuniform with $\sigma_N^2(R)\sim R^{d}$ scaling in the large-$R$ limit \cite{To06_3, Zh13}.
Specifically, we directly sample the number variance for maximally saturated RSA packings of identical hard spheres with $\phi_s\approx 0.74, 0.55,$ and 0.38 for $d = 1,2,3$, respectively \cite{To06_3, Zh13, Re58, Fe80, Co88}.
For $d=1$, we consider 9987 configurations with $N = 10^7$ particles, for $d=2$, we consider $10^4$ configurations with $N = 10^4$, and for $d=3$, we consider 100 configurations with $N = 10^4$.

\subsubsection{Poisson cluster process}
The Poisson cluster process (PCP) is a strongly clustering many-particle system with large density fluctuations on large length scales, i.e., its $S(0)$ is large, but finite, and thus is nonhyperuniform.
The PCP process is generated starting from a Poisson point process with a number density $\rho_P$ \cite{La17}.
Each of these points is treated as the center of a cluster of points where the number of points in each cluster is chosen from a Poisson distribution with mean $c$.
In this work, the positions of the points around the center are chosen from an isotropic Gaussian distribution with standard deviation $r_0$, which can be regarded as the characteristic length scale of a single cluster.
This model is also known as a modified Thomas process, which is an example of a Neyman-Scott process \cite{Il08, Ch13}.
We note that, because of the strong clustering in this system, $\rho^{1/d}$ overestimates the mean nearest-neighbor distance.
The pair statistics for this process in $\mathbb{R}^d$ are given by \cite{Il08}
\begin{equation}
    g_2(r) = 1+\frac{c}{\rho(4\pi r_0^2)^{d/2}}e^{-\frac{r^2}{4r_0^2}}
\end{equation}
and
\begin{equation}\label{eq:SkPCP}
    S(k)=1+ce^{-k^2r_0^2}.
\end{equation}
At $k=0$, $S(0) =  1 +c$, so this point pattern is nonhyperuniform and super-Poissonian [$S(0)>1$] with $\sigma_N^2(R)\sim R^{d}$ scaling in the large-$R$ limit.
Here, we use Eqs. (\ref{eq:SkPCP}) and (\ref{eq:nv}) to compute the number variance for $d = 1,2$ and 3 with $r_0\rho^{1/d}=1$, $\rho = \rho_p c =1$, $\rho_p=0.1$, and $c = 10$.

\subsubsection{Randomly vacated lattices}

Kim and Torquato \cite{Ki18} showed that spatially uncorrelated point vacancies in a many-particle system have the following effect on $S(\mathbf{k})$:
\begin{equation}\label{eq:skRVL}
    S(\mathbf{k}) = p + (1 - p)S_0(\mathbf{k})
\end{equation}
where $p$ is the fraction of bodies vacated and $S_0(\mathbf{k})$ is the structure factor of the system without vacancies.
From Eq. (\ref{eq:skRVL}), one can see that $S(k) \geq p$, where the equality holds if the system without vacancies is hyperuniform.
Thus, these systems are nonhyperuniform with $\sigma_N^2(R)\sim R^{d}$ scaling in the large-$R$ limit.
Subsequently, they showed that the number variance can be written as
\begin{equation}\label{eq:nvRVL}
    \sigma_N^2(R;p) = (1-p)p\rho v_1(R)+(1-p)^2\sigma_N^2(R;0).
\end{equation}
Here, we consider randomly vacated lattices (RVL), or lattices in which some fraction $p$ of the particles are removed at random.
In particular, we use Eqs. (\ref{eq:nvRVL}) and (\ref{eq:sklattice}) to compute the number variance for the randomly vacated hypercubic lattice $\mathbb{Z}^d_p$ for $d = 1,2$ and 3 with $p = 0.01, 0.02$, and $0.05$.

\subsubsection{Hyposurficial point process}
Hyposurficial systems are a special class of nonhyperuniform state that behave like the Poisson point process in their large-scale number variance because they lack a ``surface-area'' term proportional to $R^{d-1}$ in the large-$R$ expansion of the number variance, i.e., $\sigma_N^2(R)\sim AR^d + o(R^{d-1})$ as $R\rightarrow \infty$, where $A>0$.
Hyposurficial pair statistics obey the following sum rule \cite{To03}:
\begin{equation}
    \int_0^{\infty}r^dh(r)dr = 0,
\end{equation}
which implies that they generally contain both negative and positive correlations.
Here, we use the 3D designer hyposurficial pair statistics from Ref. \citenum{Wa24}, given by
\begin{equation}
    g_2(r) = 1 + \frac{e^{-r^*}}{4\pi}-\frac{e^{-r^*}\textrm{sin}(r^*)}{r^*}
\end{equation}
where $r^* = (4\pi)^{1/3}r$ and
\begin{equation}
    S(k) = \frac{6k^{*8}+12k^{*6}+19k^{*4}+24k^{*2}+16}{6(k^{*2}+1)^2(k^{*2}-2k^{*}+2)(k^{*2}+2k^{*}+2)}
\end{equation}
where $k^* = (4\pi)^{-1/3}k$ with Eq. (\ref{eq:nv}) to compute the number variance.
This hyposurficial point pattern is standard nonhyperuniform with $\sigma_N^2(R)\sim R^{d}$ scaling in the large-$R$ limit.

\subsubsection{Hyperplane intersection process}
The hyperplane intersection process (HIP) is a hyperfluctuating many-particle system \cite{To18}, meaning its number variance scales faster than the volume of the observation window, equivalently, $\lim_{k\rightarrow0}S(k)=\infty$ \cite{He06, Kl19}.
Such a system is also referred to as antihyperuniform because $S(k)$ diverges to $+\infty$ as $k \rightarrow 0$, which is diametrically opposed to hyperuniformity [i.e., $S(k)\rightarrow0$ as $k\rightarrow0$].
This super-Poissonian and antihyperuniform many-particle system is defined as the set of intersection points of randomly and independently distributed hyperplanes \cite{Ch13, Sc08}.
The pair correlation function for this process in $\mathbb{R}^d$ for any $d>2$ is given by \cite{He06}
\begin{equation}\label{eq:g2HIP}
    g_2(r) = 1+\sum_{k=1}^{d-1}\binom{d-1}{k}\left(\frac{\omega_{d-k}}{\omega_d
    }\right)^2\left(\frac{d\omega_d}{\omega_{d-1}}\right)^k\frac{1}{(sr)^k}
\end{equation}
where $s$ is the specific surface (i.e., the expected surface area per unit volume) of the hyperplane process and $\omega_d$ is the volume of a $d$-dimensional unit sphere.
The number density of this process is given by
\begin{equation}
    \rho = \omega_d\left(\frac{\omega_{d-1}}{d\omega_d}\right)^ds^d.
\end{equation}
For any $d$, this process has $\alpha = -1$ and large-$R$ scaling $\sigma^2_N(R)\sim R^{2d-1}$ and thus this process cannot exist for $d = 1$ \cite{Kl21}.
Here, we use Eqs. (\ref{eq:g2HIP}) and (\ref{eq:nv}) to compute the number variance for $d=2$ and 3.
We also note that, like PCP, HIP has a high degree of clustering and thus $\rho^{1/d}$ overestimates the mean nearest-neighbor distance.

\subsubsection{Hard-core antihyperuniform process}
We also examine a designer antihyperuniform process from Ref. \citenum{Wa24} that has a hard-core interaction (HC-AHU).
As described in Sec. IIB7, this means that $S(k)$ will diverge as $k\rightarrow0$, opposite to the behavior of a hyperuniform structure factor \cite{To18}.
Specifically, the pair correlation function for this system is given by:
\begin{equation}\label{eq:AHUg2}
    g_2(r) = \Theta(r/D - 1)\left(\frac{A}{(r/D)^{d-1/2}}+1\right),
\end{equation}
where $A$ is a positive constant and $D$ is the diameter of the hard-core interaction, both chosen such that the system is numerically realizable.
Such a system has $\alpha = -1/2$ and thus is super-Poissonian and antihyperuniform with $\sigma_N^2(R)\sim R^{d+(1/2)}$ scaling in the large-$R$ limit.
This pair correlation function mimics one that corresponds to a fluid at a thermodynamic critical point, though notably not one in the Ising universality class \cite{Bi92}.
Here, we use Eq. (\ref{eq:AHUg2}) with $A = 0.1$ and $\phi = 0.1$ and Eq. (\ref{eq:nv}) for $d = 1,2$ and 3 to compute the number variance.

\begin{figure}
    \centering
    \subfigure[]{\includegraphics[height=0.36\textwidth]{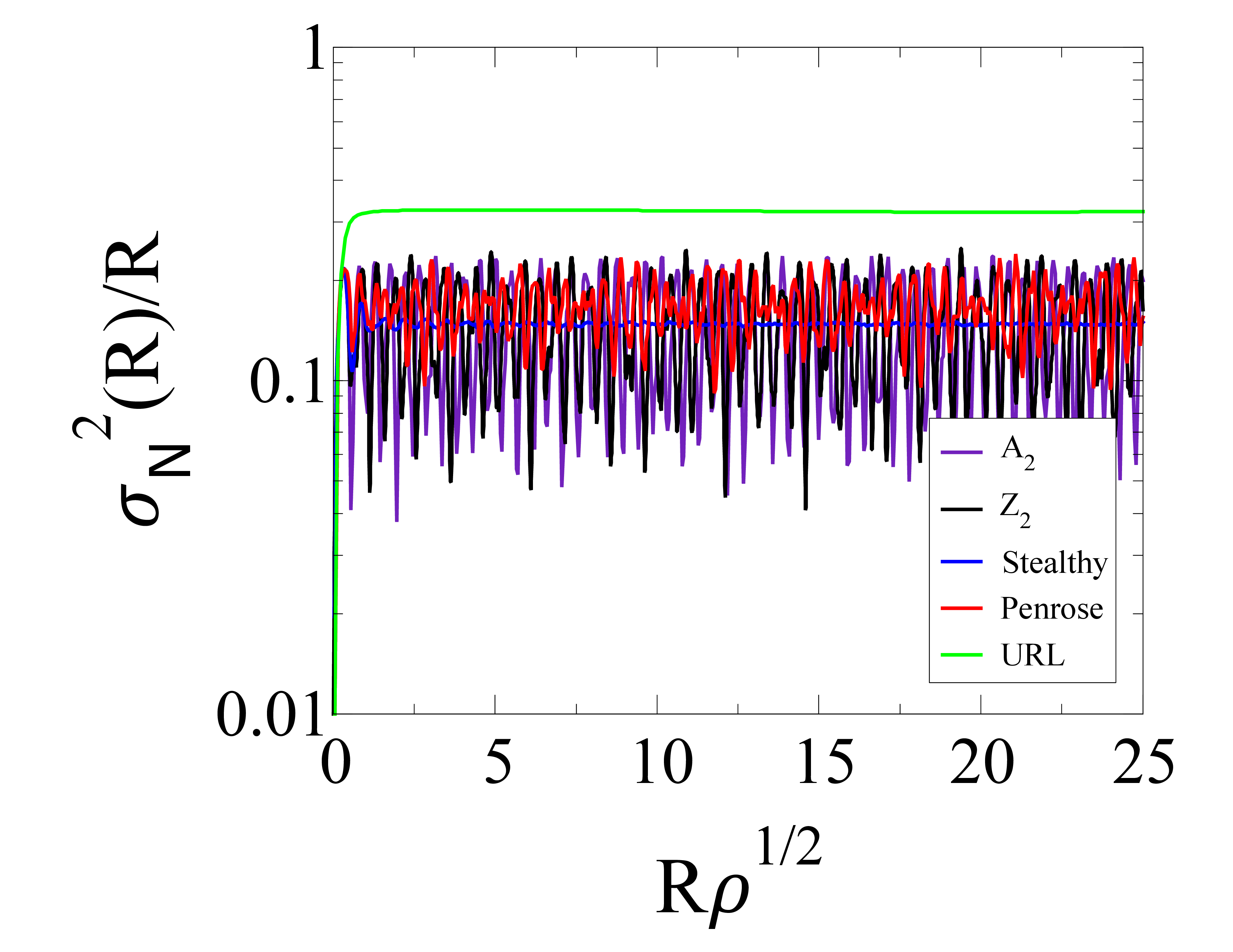} }
    \subfigure[]{\includegraphics[height=0.36\textwidth]{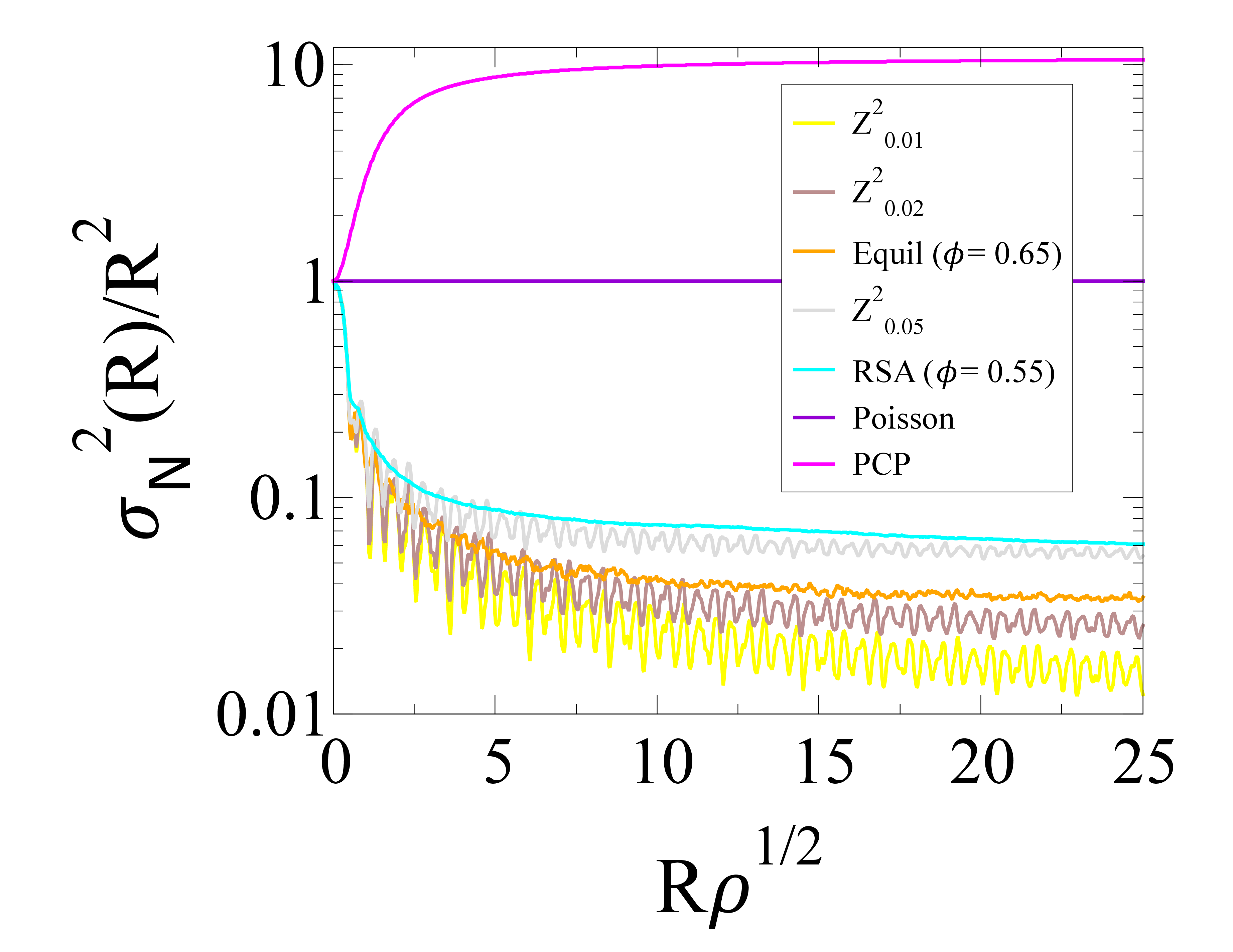} }
    \subfigure[]{\includegraphics[height=0.36\textwidth]{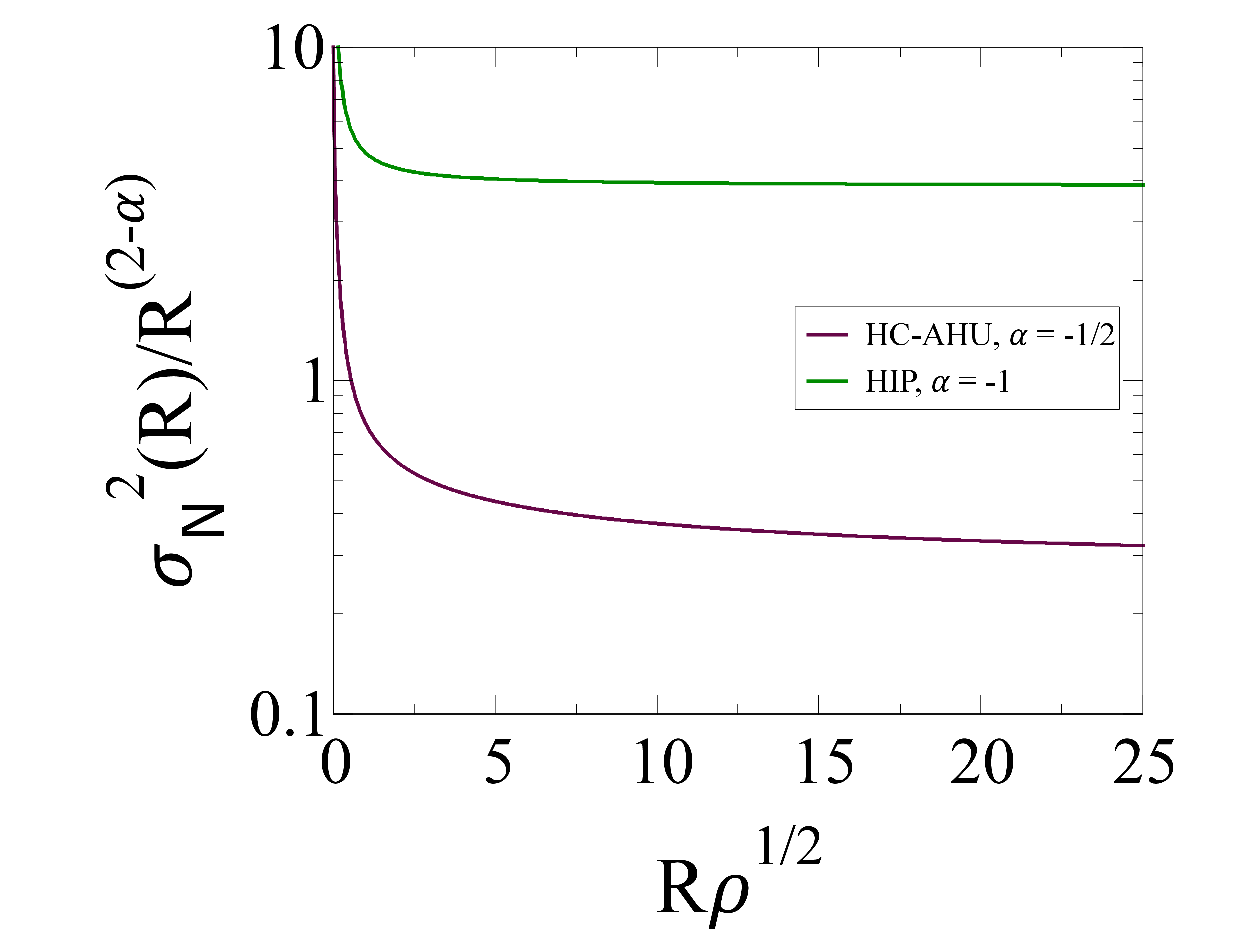} }
    \caption{Number variance curves $\sigma_N^2(R)$ for the 2D (a) hyperuniform, (b) nonhyperuniform, and (c) antihyperuniform models described in Sec. \ref{sec:models} divided by their large-$R$ scaling as a function of scaled window radius $R\rho^{1/2}$.}
    \label{fig:2Dscaling}
\end{figure}

\begin{figure*}[!t]
    \centering
    \includegraphics[width=0.79\textwidth]{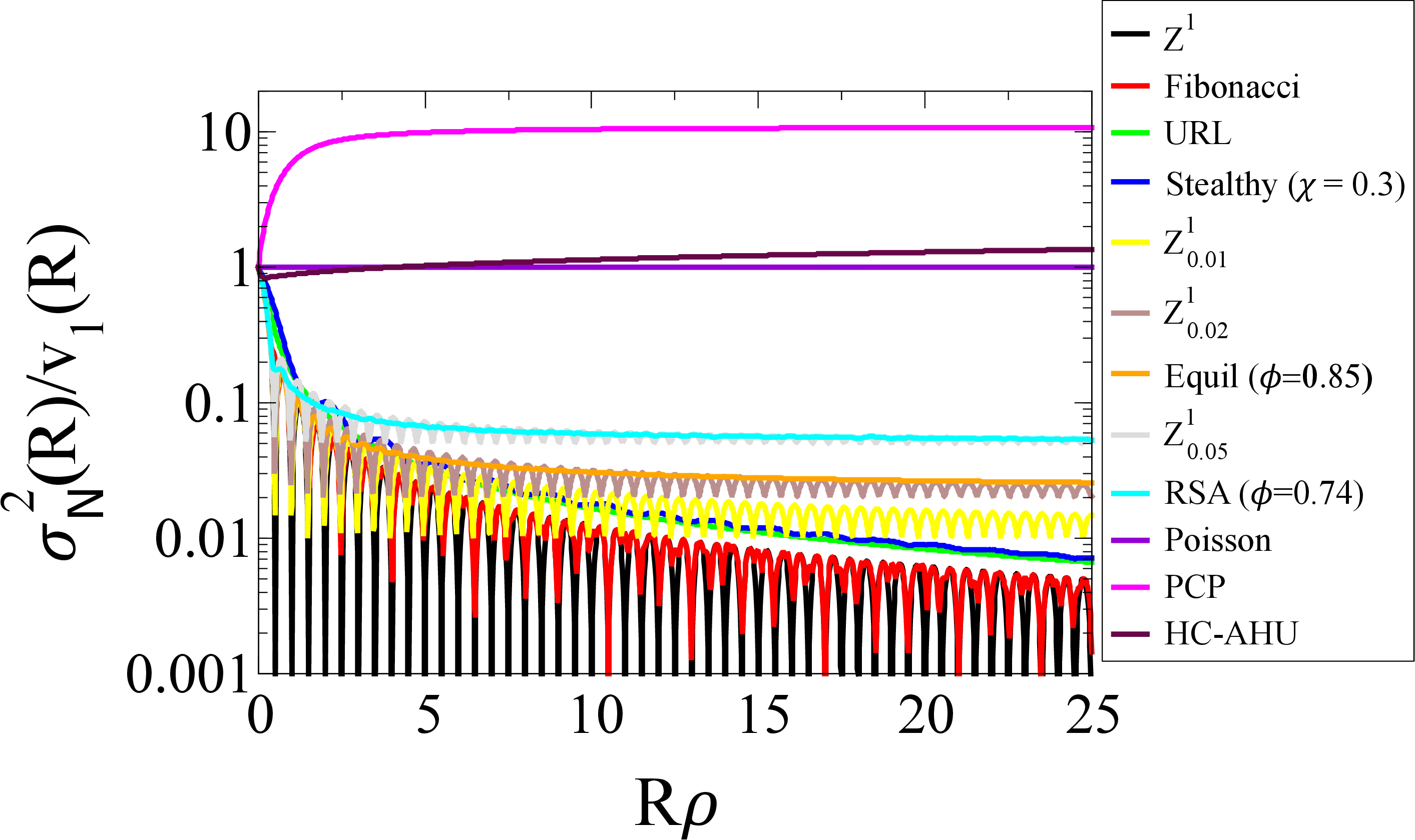}
    \caption{Comparison of the local number variance scaled by the observation window volume $\sigma_N^2(R)/v_1(R)$ versus the dimensionless window radius $R\rho$ for 1D models. For any particular value of $R$, the lower (higher) the value of $\sigma_N^2(R)$, the lower (higher) the number density fluctuations, which is a measure of a greater degree of order (disorder).}
    \label{fig:1dnv}
\end{figure*}

\begin{figure*}[!t]
    \centering
    \includegraphics[width=0.79\textwidth]{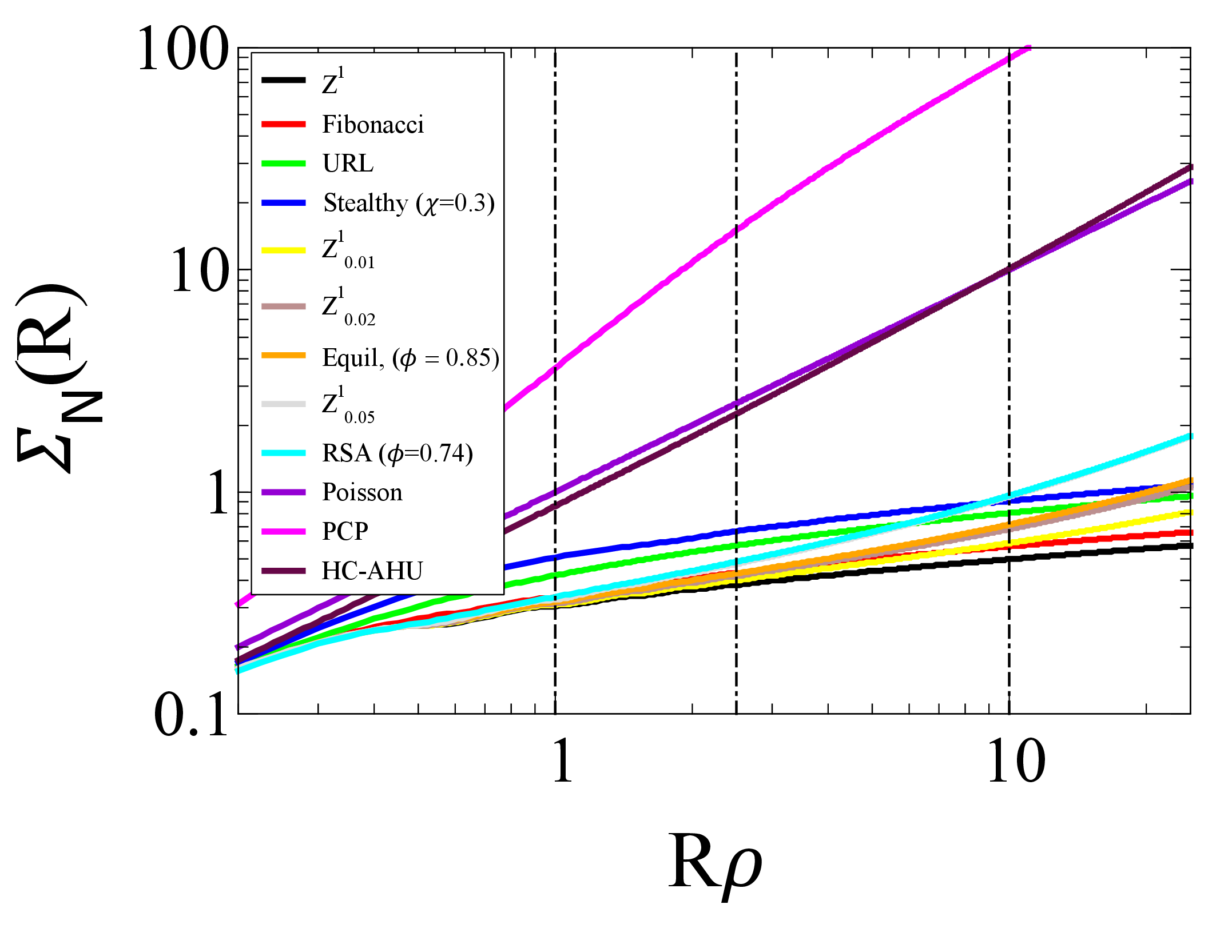}
    \caption{Comparison of the integrated local number variance $\Sigma_N(R)$ versus the dimensionless distance $R\rho$ for 1D models. For any particular value of $R$, the lower (higher) the value of $\Sigma_N(R)$, the lower (higher) the number density fluctuations, which is a measure of a greater degree of order (disorder). The dashed vertical lines denote the small ($R\rho=1$), intermediate ($R\rho=2.5$) and large ($R\rho=10$) length scales. Note that, at this scale, the cyan saturated RSA and gray $\mathbb{Z}^1_{0.05}$ curves nearly coincide.}
    \label{fig:1dnvint}
\end{figure*}

\begin{table*}[!h]
\caption{\textcolor{black}{Compiled asymptotic characteristics for certain class I hyperuniform models including the $A_d^*$, $A_d$, and $\mathbb{Z}^d$ lattices, disordered stealthy hyperuniform point patterns (Stealthy), Fibonnacci and Penrose quasicrystals, and uniformly randomized lattices (URL). Also included are certain antihyperuniform models including the hyperplane intersection process (HIP) and hard-core antihyperuniform process (HC-AHU) as well as certain standard nonhyperuniform models including the Poisson cluster process (PCP), Poisson process, hyposurficial point process, random sequential addition packings (RSA), randomly vacated lattices ($\mathbb{Z}^d_p$), and equilibrium hard rod, disk, and sphere packings (Equil.). Specifically, tabulated here are the scaling exponent $\alpha$, volume coefficient $A$, surface area coefficient $B$, and the ratio $B/A$ for the aforementioned models across the first three space dimensions. Values of $A$ and $B$ are determined analytically where possible using Eqs. (\ref{eq:A}) and (\ref{eq:B}), respectively, and by fitting the number variance using a polynomial with the form of Eq. (\ref{eq:nv_asy}) otherwise. Stealthy hyperuniform models can be roughly regarded to have a scaling exponent $\alpha=+\infty$, which strictly speaking is not mathematically precise since there is no such limiting process.}}
\begin{tabular}{lllll}
\hline\hline
Model                 &$\alpha$ & A      & B       & B/A    \\
\hline
2D HIP& -1 &$+\infty$& - & - \\
3D HIP & -1 &$+\infty$& - & - \\
\hline
1D HC-AHU & -1/2 &$+\infty$& - & - \\
2D HC-AHU & -1/2 &$+\infty$& - & - \\
3D HC-AHU & -1/2 &$+\infty$& - & - \\

\hline
1D PCP                &0 & 11     & -       & -      \\
2D PCP                &0 & 11     & -       & -      \\
3D PCP                &0 & 11     & -       & -      \\
\hline
1D Poisson            &0 & 1      & 0       & 0      \\
2D Poisson            &0 & 1      & 0       & 0      \\
3D Poisson            &0 & 1      & 0       & 0      \\
\hline
3D Hyposurficial      &0 & 2/3    & 0       & 0      \\
\hline
1D RSA ($\phi = 0.74$)              & 0& 0.0507 & 0.08139 & 1.6053 \\
2D RSA ($\phi = 0.55$)              & 0& 0.0586 & 0.1447  & 2.4693 \\
3D RSA ($\phi = 0.38$)              & 0& 0.0511 & 0.2731  & 5.348  \\
\hline
$\mathbb{Z}^1_{0.05}$   &0 & 0.05   & 0.07521 & 1.504  \\
$\mathbb{Z}^2_{0.05}$   &0 & 0.05   & 0.1349  & 2.698  \\
$\mathbb{Z}^3_{0.05}$   &0 & 0.05   & 0.1867  & 3.734  \\
\hline
$\mathbb{Z}^1_{0.02}$   &0 & 0.02   & 0.08003 & 4.002  \\
$\mathbb{Z}^2_{0.02}$   &0 & 0.02   & 0.1413  & 7.065  \\
$\mathbb{Z}^3_{0.02}$   &0 & 0.02   & 0.1946  & 9.73   \\
\hline
1D Equil.  ($\phi = 0.85$)  & 0& 0.0225 & 0.08245 & 3.664  \\
2D Equil.  ($\phi = 0.65$)  & 0& 0.0236 & 0.1590  & 6.737  \\
3D Equil.  ($\phi = 0.48$)  &0 & 0.019  & 0.1990  & 10.47  \\
\hline
$\mathbb{Z}^1_{0.01}$   & 0& 0.01   & 0.08167 & 8.1675 \\
$\mathbb{Z}^2_{0.01}$   & 0& 0.01   & 0.1435  & 14.35  \\
$\mathbb{Z}^3_{0.01}$   & 0& 0.01   & 0.1973  & 19.73  \\
\hline
$A^*_3$&$+\infty$&0&0.1930&-\\
\hline
$A_2$&$+\infty$&0&0.1434&-\\
$A_3$&$+\infty$&0&0.1932&-\\
\hline
$\mathbb{Z}^1$  &  $+\infty$& 0  & 0.08$\overline{33}$ & - \\
$\mathbb{Z}^2$  & $+\infty$ & 0   & 0.1457  & -  \\
$\mathbb{Z}^3$  &  $+\infty$& 0   & 0.1999  & -  \\
\hline
1D Stealthy $\chi=0.30$&$+\infty$ &0&0.1787&-\\
2D Stealthy $\chi=0.49$&$+\infty$ &0&0.1485&-\\
3D Stealthy $\chi=0.49$& $+\infty$&0&0.1996&-\\
\hline
Fibonacci  & 3& 0 & 0.1006 & - \\
Penrose  & 6& 0 & 0.1694 & - \\
\hline
1D URL&2&0&0.1$\overline{666}$&-\\
2D URL&2&0&0.3250&-\\
3D URL&2&0&0.4950&-\\
\hline\hline
\label{tab:AB}
\end{tabular}
\end{table*}

\begin{figure*}[!t]
    \centering
    \includegraphics[width=0.75\textwidth]{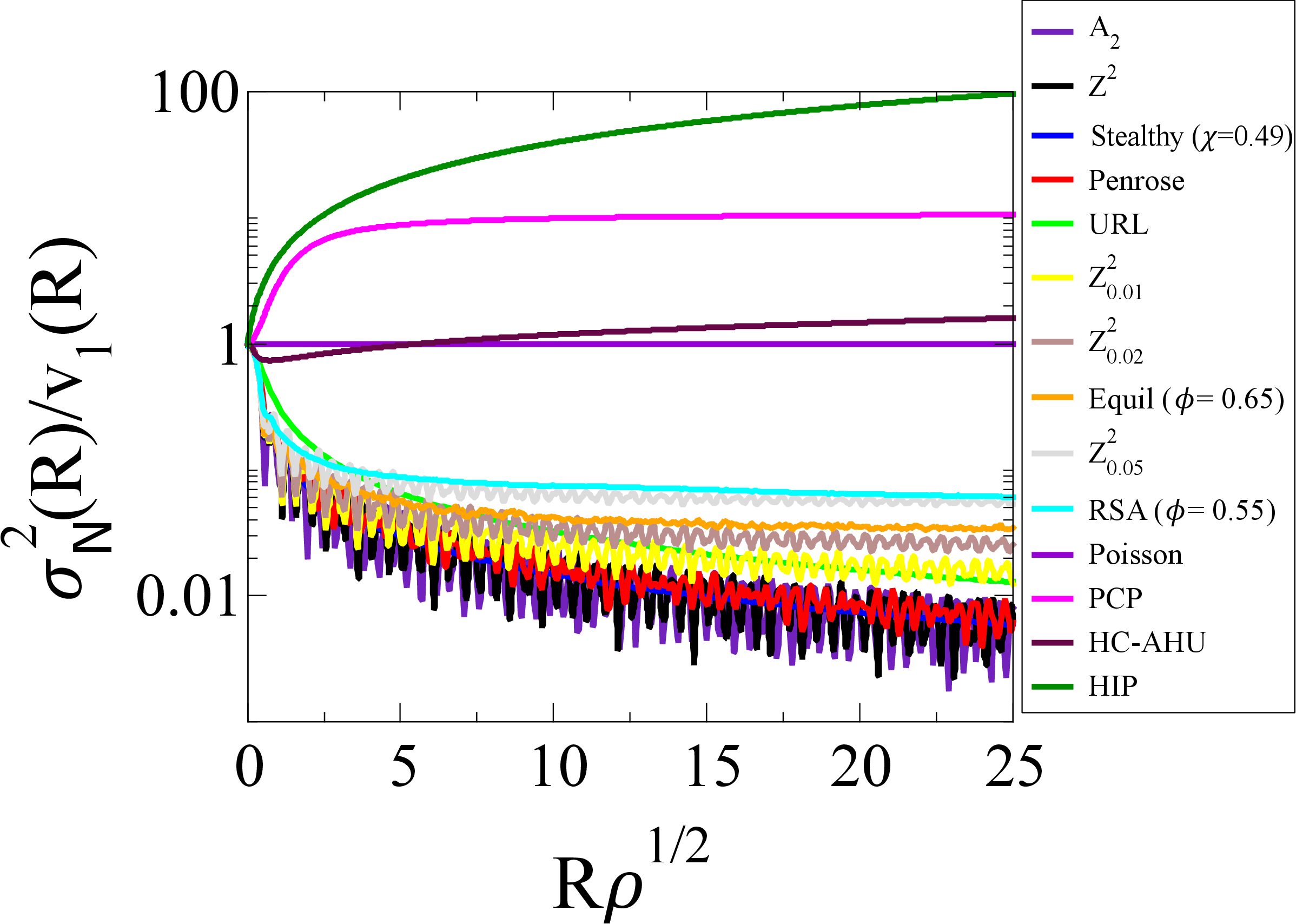}
    \caption{Comparison of the local number variance scaled by the observation window volume $\sigma_N^2(R)/v_1(R)$ versus the dimensionless window radius $R\rho^{1/2}$ for 2D models. For any particular value of $R$, the lower (higher) the value of $\sigma_N^2(R)$, the lower (higher) the number density fluctuations, which is a measure of a greater degree of order (disorder).}
    \label{fig:2dnv}
\end{figure*}
\begin{figure*}[!t]
    \centering
    \includegraphics[width=0.65\textwidth]{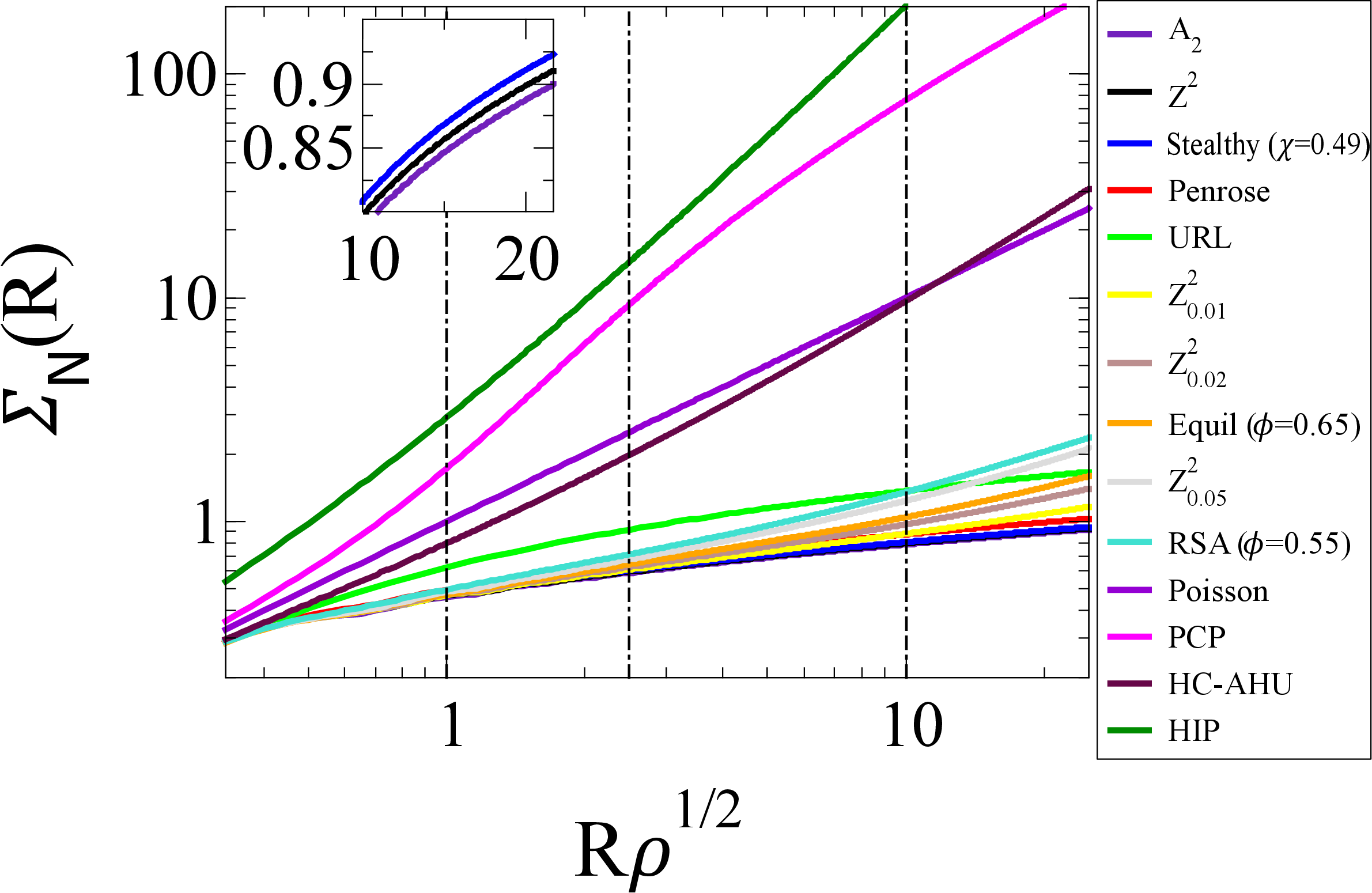}
    \caption{Comparison of the integrated local number variance $\Sigma_N(R)$ versus the dimensionless distance $R\rho^{1/2}$ for 2D models. For any particular value of $R$, the lower (higher) the value of $\Sigma_N(R)$, the lower (higher) the number density fluctuations, which is a measure of a greater degree of order (disorder). The dashed vertical lines denote the small ($R\rho^{1/2}=1$), intermediate ($R\rho^{1/2}=2.5$) and large ($R\rho^{1/2}=10$) length scales. The inset shows the large-$R$ behavior of the disordered stealthy hyperuniform, $\mathbb{Z}^2$ lattice, and $A_2$ lattice.}
    \label{fig:2dnvint}

\end{figure*}

\section{Results}

\subsection{Design of Number-Variance Based Order Metrics}

Here, we propose the use of the scaled local number variance $\sigma_N^2(R)/v_1(R)$ at sampling window radius $R$ and an integral measure derived from it $\Sigma_N(R_i,R_j)$, where $R_{i,j}$ are two prescribed length scales, as order metrics for antihyperuniform, nonhyperuniform, disordered, and ordered hyperuniform systems across length scales by tracking their behavior as a function of $R$.
The number variance is computed analytically where possible, and computed numerically via the direct sampling method described in the Appendix otherwise.
Immediately below, we justify our choice of $v_1(R)$ window-volume scaling and define $\Sigma_N(R_i,R_j)$.

Figure \ref{fig:2Dscaling} shows $\sigma_N^2(R)$ for each 2D model described in Sec. \ref{sec:models} divided by their respective large-$R$ scalings.
In Fig. \ref{fig:2Dscaling}(a), which depicts number variance curves for 2D hyperuniform systems, each curve rapidly plateaus indicating that they attain, on average, their large-$R$ scaling at small length scales, which was also shown by Torquato and Stillinger \cite{To03}.
For standard nonhyperuniform and antihyperuniform systems, the length scale at which the large-$R$ nonhyperuniform scaling sets in can vary widely across models.
In the case of disordered nearly hyperuniform point patterns, the ratio $B/A$ 
enables one to ascertain hyperuniform and nonhyperuniform distance-scaling regimes of the number variance $\sigma_N^2(R)$ as a function of $R$ as well as the corresponding crossover distance between the hyperuniform and nonhyperuniform regimes \cite{To21}.
In Table \ref{tab:AB}, we compile values of the \textcolor{black}{scaling exponent $\alpha$,} volume coefficient $A$, surface area coefficient $B$, \textcolor{black}{and the ratio $B/A$} for \textcolor{black}{the models considered in this work}.
By comparing the values of $B$ in Table \ref{tab:AB} with the large-$R$ behavior in Fig. \ref{fig:2Dscaling} (a) and the values of $A$ in Table \ref{tab:AB} with the large-$R$ behavior in Fig. \ref{fig:2Dscaling} (b), one can see that $B$ can be used to rank order the large-$R$ degree of order/disorder in hyperuniform systems, while $A$ can likewise be used for the standard nonhyperuniform systems.
Clearly, across models in a given space dimension, there is a wide range of $B/A$ values and, for a given model across space dimensions, this ratio increases.
For the RVL, the defect concentration $p = A$, so when $p$ is made small $B/A$ becomes large. Thus, when $p$ is sufficiently small, RVLs can have hyperuniform $\sigma_N^2(R)$ scaling up to several orders of magnitude in $R\rho^{1/d}$, while having nonhyperuniform scaling on larger length scales.
The presence of well-defined hyperuniform \textit{and} nonhyperuniform variance scaling regimes makes the RVL system an excellent nontrivial example of length scale dependent order in a system and clearly exemplifies why one must specify a length scale when assessing the order/disorder within a system.
In particular, a global assessment of order should group the RVL with other nonhyperuniform systems, meaning one would miss, e.g., that the $\mathbb{Z}_{0.01}^2$ number variance curve in Fig. \ref{fig:2dnv} has similar oscillations and scaling behavior to the ordered hyperuniform systems (Penrose, $\mathbb{Z}^2,\;A_2$) up to approximately the $B/A$ value for $\mathbb{Z}_{0.01}^2$ listed in Table \ref{tab:AB} (14.35).
The treatment of the number variance described above only allows one to compare large scale order within a particular asymptotic scaling class (see Sec. \ref{sec:HU}) or, in the case of nearly nonhyperuniform systems, the transition length scale at which large-$R$ scaling sets in.

We choose order metrics here based on the local number variance scaled by $v_1(R)$ for \textit{all} systems, as opposed to scaling by their respective large-$R$ behaviors, so we can fairly compare the degree of order/disorder \textit{across} length scales.
For example, at small $R$, one would expect systems with hard-core interactions to have very similar short-range order.
However, comparing the small-$R$ regimes in Fig. \ref{fig:2Dscaling} (a) and Fig. \ref{fig:2Dscaling} (b), we can see that the hard-core processes [e.g., $A_2$ and $\mathbb{Z}^2$ in Fig. \ref{fig:2Dscaling} (a); RSA and equilibrium hard disks in  \ref{fig:2Dscaling} (b)] have extremely different small-$R$ behaviors.
Using $v_1(R)$, all curves will have $\sigma_N^2(0)/v_1(0)=1$, enabling us to compare the small-$R$ behavior of models across the different large-$R$ scaling classes.
This scaling choice is further motivated by the success of other variance-based order metrics that use a $v_1(R)$ scaling \cite{To22, Sk24}.
Moreover, with this window-volume scaling, $\sigma_N^2(R)/v_1(R)$ will approach $+\infty$ as $R\rightarrow\infty$ for antihyperuniform systems, $S(0)$ (positive and bounded) as $R\rightarrow\infty$ for standard nonhyperuniform systems, and 0 as $R\rightarrow\infty$ for hyperuniform systems.
Thus, with the $v_1(R)$ scaling, these metrics still maintain the ability to assess the degree of large-$R$ order/disorder across models.

To reduce the effect of fluctuations that occur for all $R$ in the number variance for systems with long-ranged order (e.g., lattices and RVL) or from noise in numerically sampled number variance curves, we additionally consider an integral measure derived from $\sigma_N^2(R)/v_1(R)$
\begin{equation}
    \Sigma_N(R_i,R_j) = \int_{R_i}^{R_j}\sigma_N^2(\ell)/v_1(\ell)d\ell
\end{equation}
where $R_i<R_j$ are two length scales and $\Sigma_N(R)$, i.e., with a single argument, implies $R_i = 0$.
In addition to reducing fluctuations, this integral metric enables us to characterize the degree of order/disorder in a point pattern over a prescribed range of length scales.
\textcolor{black}{Here, we focus on the single-argument function $\Sigma_N(R)$, which provides an integrated measure of the degree of order/disorder in a system up to a length scale $R$ and contrasts with $\sigma_N^2(R)/v_1(R)$, which measures the degree of order/disorder {\it only} at the length scale $R$.}
As above, $\Sigma_N(R)$ will have three large-$R$ scaling types: superlinear for antihyperuniform systems, linear for nonhyperuniform systems, and sublinear for hyperuniform systems.
\textcolor{black}{In Sec. V, we discuss the utility of the two-argument $\Sigma_N(R_i,R_j)$.}

\subsection{Application of Number Variance-Based Order Metrics for Models in $d=1,2,3$}
In the following subsections, we present results for the scaled local number variance $\sigma_N^2(R)/v_1(R)$ obtained analytically when possible via Eq. (\ref{eq:nv}) or via direct sampling (see the Appendix) and the related integrated measure $\Sigma_N(R_i,R_j)$ for the 1D, 2D, and 3D models discussed in Sec. \ref{sec:models}.
To compare different models in a particular space dimension, all distances are scaled by $\rho^{1/d}$, which is a reasonable approximation of the mean nearest-neighbor distance between particles in a system (see Sec. \ref{sec:models} for exceptions).

\subsubsection{1D Models}
Figure \ref{fig:1dnv} shows the scaled number variance $\sigma_N^2(R)/v_1(R)$ as a function of the scaled window radius $R\rho$ for each of the 1D models described in Sec. \ref{sec:models}.
Changes in the $\sigma_N^2(R)/v_1(R)$ curves as a function of $R$ indicate it is sensitive to changes in the spatial correlations between pairs of particles at the corresponding length scales. In addition, intersections between these curves show that the relative rankings of order/disorder of these systems can change across length scales.
One can relate the behavior of $\sigma_N^2(R)/v_1(R)$ at short, intermediate, and large length scales to various characteristics of their pair statistics, for example, the greatest degree of short-range order [fastest decay of $\sigma_N^2(R)/v_1(R)$] is attained by systems whose particles have exclusion regions, i.e. that have $h(r)=-1$ for $0<r<r'$, which can correspond to hard-particle systems or those with strong repulsive potentials. 
Slower small-$R$ decay (or growth) is associated with weaker short-range anticorrelations, i.e., $h(r) > -1$ (or short-range correlations, i.e., $h(r) > 0$) in the vicinity of the origin.
Oscillations at small $R$ reflect local translational order and the extent to which these oscillations extend into intermediate length scales corresponds to longer-ranged translational order. Oscillations at all $R$ indicate long range order (i.e., Bragg peaks in $S(k)$).
The separation of the curves into the three scaling types reflects their respective degrees of large-$R$ order: $\sigma_N^2(R)/v_1(R)$ curves for antihyperuniform systems continue to increase toward $+\infty$ at large $R$, $\sigma_N^2(R)/v_1(R)$ curves for nonhyperuniform systems plateau to their corresponding $S(0)$ (equivalently, $A$) values at large $R$, and $\sigma_N^2(R)/v_1(R)$ curves hyperuniform systems continue to decrease toward 0 at large $R$. Comparing the small- and large-$R$ behavior of, e.g., RSA and URL, one can see that the (non)hyperuniformity of a system (i.e., its long-ranged order) is not necessarily related to its degree small-$R$ order/disorder.

Figure \ref{fig:1dnvint} shows $\Sigma_N(R)$ as a function of $R\rho$ for each of the systems considered in Fig. \ref{fig:1dnv}.
Here, and in subsequent sections, we consider $R\rho^{1/d}=1$ (approximately 2 mean nearest-neighbor distances) to be a ``short'' length scale, $R\rho^{1/d}=2.5$ (approximately 5 mean nearest-neighbor distances) to be an ``intermediate'' length scale, and $R\rho^{1/d}=10$ (approximately 20 mean nearest-neighbor distances) to be a ``long'' length scale (see vertical lines in Fig. \ref{fig:1dnvint}).
In Fig. \ref{fig:1dnvint}, it is evident that the rank-order of the 1D point patterns is different at each of the vertical lines.
As described above, the rank-order of the systems at small length scales is directly related to the small-$r$ behavior of their respective $h(r)$. A notable exception to this behavior is the HC-AHU system, whose hard-core has a radial extent significantly smaller than the other hard-core systems and is significantly more disordered by comparison.
At intermediate length scales, the hard-core nonhyperuniform systems become more disordered relative to the hard-core hyperuniform systems, but are still more ordered than the non-hard-core hyperuniform systems.
At large length scales, one can observe the three large-$R$ scaling classes described above: nonhyperuniform systems scale linearly, hyperuniform systems have sublinear growth, and antihyperuniform systems have superlinear growth. In addition, we observe crossover between nonhyperuniform hard-core system curves and non-hard-core hyperuniform system curves, as well as between the HC-AHU curve and the nonhyperuniform Poisson curve.
As described in Sec. IV A, the large-$R$ ($R\rho^{1/d}=10$) rank order here for the hyperuniform and nonhyperuniform systems match the $A$ and $B$ rank ordering, respectively, and the different values of $B/A$ for different hard-core nonhyperuniform systems explain why these curves separate from the hard-core hyperuniform systems at different length scales (cf. Table \ref{tab:AB}).

\subsubsection{2D Models}
Figure \ref{fig:2dnv} shows the scaled number variance $\sigma_N^2(R)/v_1(R)$ as a function of the scaled window radius $R\rho^{1/2}$ for each of the 2D models described in Sec. \ref{sec:models}.
The general qualitative statements regarding the scaling classes and behaviors of the scaled number variance curves corresponding to structural motifs shown in the pair statistics of a given system follow from the $d = 1$ discussion.
With the addition of the antihyperuniform HIP system, one can observe clear, qualitative differences across length scales between the $\sigma_N^2(R)/v_1(R)$ curves of the two antihyperuniform systems that are related to their extremely different $h(r)$ as $r\rightarrow 0$ behaviors. While HC-AHU has small-$R$ decay due to its hard core and subsequent large-$R$ growth, HIP increases extremely rapidly at small $R$ because its $h(r)$ diverges to $+\infty$ at the origin and is the most disordered system examined here across \textit{all} length scales.
Another interesting qualitative difference between the $d = 1$ and $d = 2$ models is that, unlike for $d=1$, where the number variance curve for the high-$\chi=0.3$ stealthy hyperuniform point patterns nearly matched that of the URL, the number variance curve for the high-$\chi=0.49$ stealthy hyperuniform point patterns here more closely follows those of the periodic and quasiperiodic point patterns for $d=2$. In other words, one can get closer to optimal suppression of number density fluctuations (see Ref. \cite{To10_2}) with disordered stealthy hyperuniform point patterns in $d \geq 2$ than in $d=1$. This discrepancy is related to differences in the maximum value of $\chi$ for disordered stealthy hyperuniform point patterns between $d=1$ (0.33) \cite{Fa91} and $d \geq 2$ (0.5) \cite{Zh15}.

\begin{figure*}[!t]
    \centering
    \includegraphics[width=0.65\textwidth]{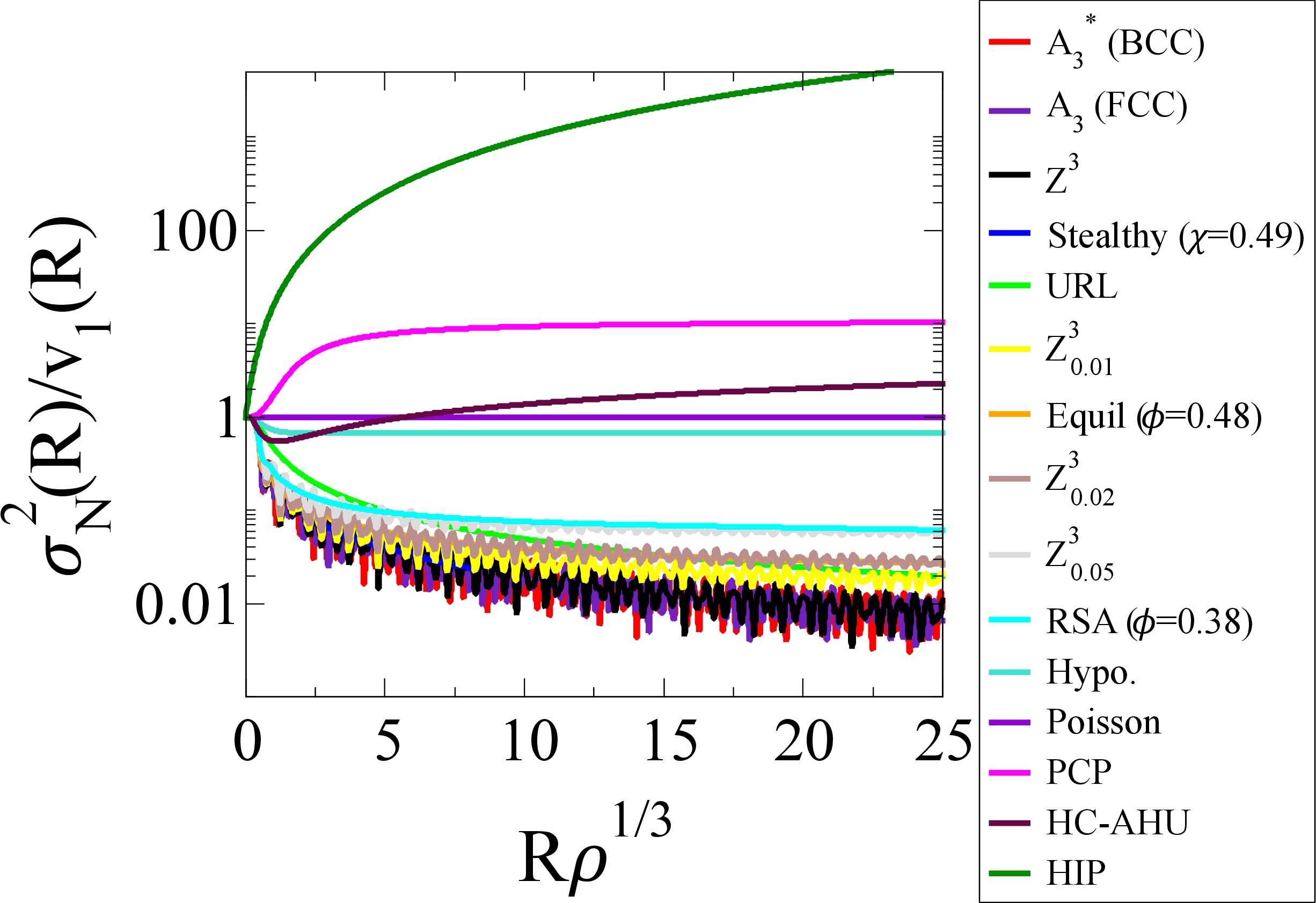}
    \caption{Comparison of the local number variance scaled by the observation window volume $\sigma_N^2(R)/v_1(R)$ versus the dimensionless window radius $R\rho^{1/3}$ for 3D models. For any particular value of $R$, the lower (higher) the value of $\sigma_N^2(R)$, the lower (higher) the number density fluctuations, which is a measure of a greater degree of order (disorder).}
    \label{fig:3dnv}

\end{figure*}
\begin{figure*}[!t]
    \centering
    \includegraphics[width=0.65\textwidth]{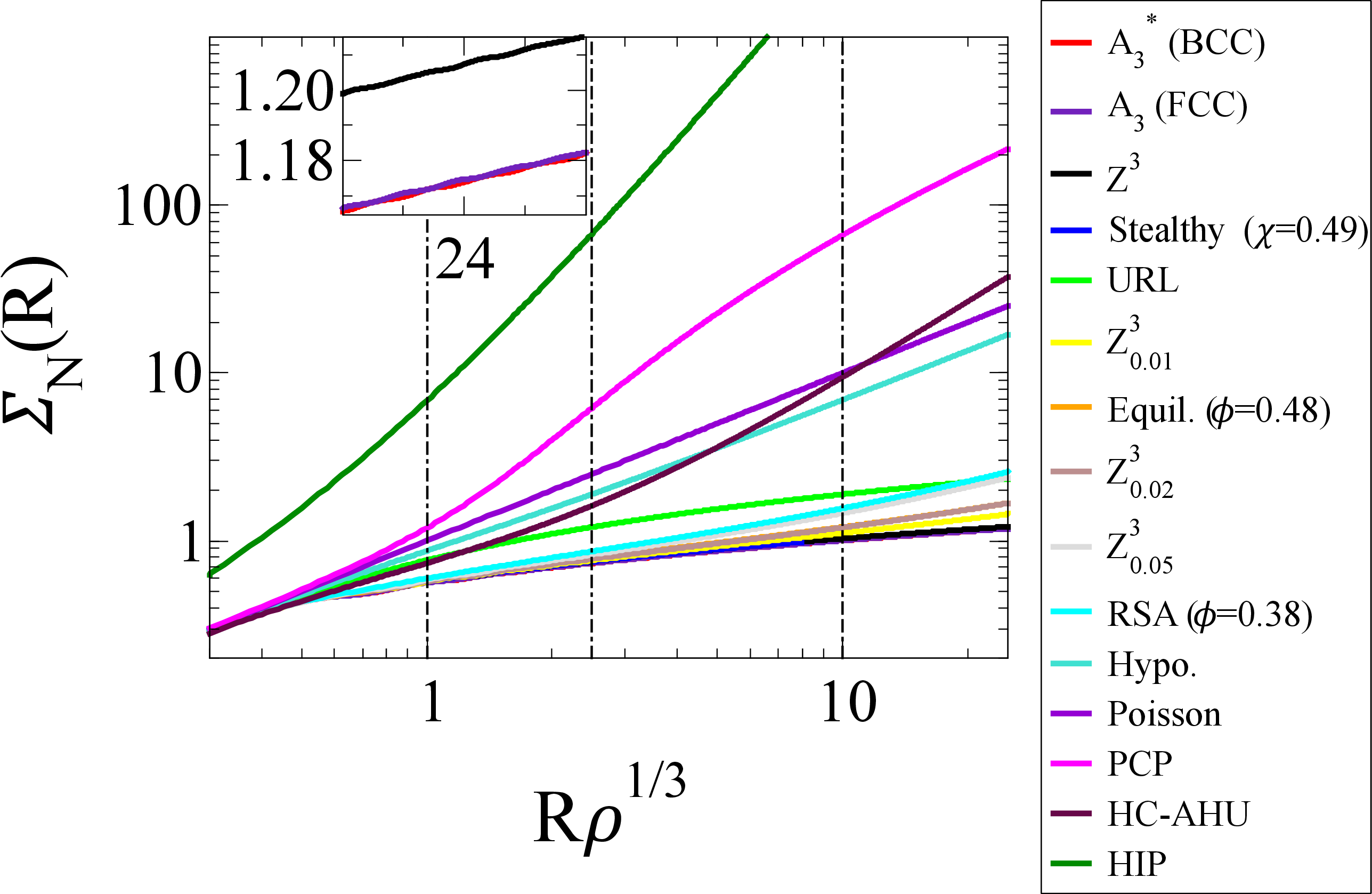}
    \caption{Comparison of the integrated local number variance $\Sigma_N(R)$ versus the dimensionless distance $R\rho^{1/3}$ for 3D models. For any particular value of $R$, the lower (higher) the value of $\Sigma_N(R)$, the lower (higher) the number density fluctuations, which is a measure of a greater degree of order (disorder). The dashed vertical lines denote the small ($R\rho^{1/3}=1$), intermediate ($R\rho^{1/3}=2.5$) and large ($R\rho^{1/3}=10$) length scales. The inset shows the large-$R$ behavior of the $\mathbb{Z}^3$, $A_3$ (fcc), and $A_3^*$ (bcc) lattices. Note that, at this scale, the orange equilibrium hard sphere and brown $\mathbb{Z}^1_{0.02}$ curves nearly coincide.}
    \label{fig:3dnvint}
\end{figure*}

Figure \ref{fig:2dnvint} shows $\Sigma_N(R)$ as a function of $R\rho^{1/2}$ for each of systems considered in Fig. \ref{fig:2dnv}.
As above, qualitative statements regarding the relative rankings of different types of point patterns follow from the discussion of $\Sigma_N(R)$ for $d = 1$.
The key differences in rankings from the $d = 1$ results include the following: at small length scales, the newly introduced HIP system is the most disordered at small length scales of all systems considered. Additionally, there is a significant increase in the order of the high-$\chi$ stealthy systems due to the emergence of real-space hard-core-like character in $d\geq 2$ (cf. Figs. S5 and S16).
We also find that for $d = 2$ $A_2$ is the most ordered structure, as opposed to $\mathbb{Z}^d$ (cf. Fig. \ref{fig:1dnvint}), consistent with the expected asymptotic result given their respective $B$ values.
Moreover, unlike for $d = 1$, we do not observe a crossover at large $R$ between the non-hard-core hyperuniform systems and hard-core nonhyperuniform systems. This difference may be a result of the small-$R$ suppression of fluctuations due to a hard core being a more dominant effect in higher space dimensions because small-$R$ behavior is weighted more strongly by $\alpha_2(r;R)$ than large-$R$ behavior in higher dimensions [cf. Eq. (\ref{a2})].
This observation is also consistent with the increasing value of $B/A$ for nonhyperuniform systems as $d$ increases \cite{To21}.

\subsubsection{3D Models}
Figure \ref{fig:3dnv} shows the scaled number variance $\sigma_N^2(R)/v_1(R)$ as a function of the scaled window radius $R\rho^{1/3}$ for each of the 3D models described in Sec. \ref{sec:models}.
As above, the general qualitative characteristics of the curves follow from the discussion for $d = 1$ and 2.
Here, we note that, across dimensions, the hard-core hyperuniform and nonhyperuniform system curves are clustered more tightly across the length scales considered in Fig. \ref{fig:3dnv} due to the crystalline point patterns becoming relatively more disordered as $d$ increases, consistent with previous findings and the decorrelation principle \cite{To06,An16}. Similarly, HIP and HC-AHU have larger number density fluctuations at large $R$ for $d = 3$ than for $d = 2$.

Figure \ref{fig:3dnvint} shows $\Sigma_N(R)$ as a function of $R\rho^{1/3}$ for each of the systems considered in Fig. \ref{fig:3dnv}.
As above, qualitative statements regarding the relative rankings of different types of point patterns follow from the discussion of $\Sigma_N(R)$ for $d = 1$ and 2.
The key differences from $d = 1$ and 2 include that, at small length scales, the fcc lattice is more ordered than the bcc lattice, which runs counter to the large-$R$ asymptotic result of bcc being more ordered (cf. the $B$ values of the fcc and bcc lattices in Table \ref{tab:AB}). Additionally, for $d = 3$ HC-AHU is more ordered than URL, which is not the case in $d = 1,2$, due to the range of its hard-core interactions increasing.
At intermediate $R$, the high-$\chi$ stealthy hyperuniform point patterns are more ordered than $\mathbb{Z}^3$, consistent with their expected asymptotically large $R$ behavior given their respective $B$ values. 
This demonstrates that ordered hyperuniform point patterns are not necessarily the \textit{most} ordered possible systems at a particular length scale, consistent with findings for two-phase media \cite{To22}. The HC-AHU system also now appears more disordered than the URL. 
In the neighborhood of intermediate values of $R$, the fcc and bcc curves intersect several times meaning one must go to much larger $R$ to match the asymptotic result of the bcc lattice minimizing $B$ in 3D (cf. Table \ref{tab:AB}) \cite{To10_2}.
Moreover, the inset shows the gap between the $\mathbb{Z}^3$ lattice and the more ordered bcc and fcc lattices at large $R$, consistent with the notion that the $\mathbb{Z}^d$ lattices become less optimal (lower densities, worse solutions to the covering and quantizer problems) as $d$ increases compared to the optimal lattices in a given space dimension \cite{To10_2}.

Across space dimensions, we find that changes in ranking between systems, e.g., URL and RSA (cf. Figs. \ref{fig:1dnvint}, \ref{fig:2dnvint}, and \ref{fig:3dnvint}), tend to occur at larger length scales as $d$ increases.
This observation is consistent with the ratio $B/A$ increasing as a function of $d$ for a particular model, which indicates that the length scale at which nonhyperuniform systems reach their asymptotic scaling increases with $d$.
In addition, the smallest $\Sigma_N(R)$ values found at a particular $R$ for $d =3$ are larger than those for $d = 2$, which are larger than those for $d = 1$, indicating that the most ordered system in a given space dimension becomes more disordered as $d$ increases, consistent with the decorrelation principle \cite{To06}.

\section{Discussion and Conclusions}
In this work, we have devised order metrics to characterize the degree of order/disorder of many-body systems across all length scales via the local number variance $\sigma_N^2(R)$.
We found that the scaled number variance $\sigma_N^2(R)/v_1(R)$ and an integral measure derived from it $\Sigma_N(R_i, R_j)$ were able to sensitively characterize the degree of order/disorder across all length scales of 41 different point patterns that span antihyperuniform, standard nonhyperuniform, disordered hyperuniform, and ordered hyperuniform kinds with varying degrees of short-, intermediate-, and long-ranged order.
Specifically, we found the degree of short-scale order is related directly to the degree of (anti)correlations in the small-$r$ regime of the $h(r)$ for a particular model; systems whose particles have exclusions regions [$h(r) = -1$ for $0 < r < r'$] have the greatest degree of small-scale order, while systems with $h(r) = +\infty$ as $r\rightarrow0$ have the greatest degree of short-scale disorder.
At intermediate length scales, we found that nonhyperuniform systems with hard-core interactions began to appear more disordered relative to the hyperuniform ones and antihyperuniform systems began to appear more disordered relative to nonhyperuniform systems.
Finally, at large length scales, the aforementioned changes at intermediate length scales become more exaggerated.
Across space dimensions, we find that a particular model becomes more disordered, consistent with the decorrelation principle \cite{To06}.
Using $\Sigma_N(R)$, we showed that the ranking of the degree of order/disorder across models changes as a function of $R$, reaffirming that it is important to assess order/disorder with respect to a specific length scale \cite{To22, Sk24}.

The differences in length scale at which the nearly hyperuniform systems (e.g., RVL, equilibrium hard rods/disks/spheres) become markedly more disordered than hard-core hyperuniform systems are closely related to the ratio $B/A$, which quantifies the length scale at which a nearly hyperuniform system transitions from hyperuniform to nonhyperuniform $\sigma_N^2(R)$ scaling.
The RVL is a particularly illuminating example because the defect concentration $p$ can be made small, resulting in large values of $B/A$. Thus, for sufficiently small $p$, RVLs can have hyperuniform $\sigma_N^2(R)$ scaling up to several orders of magnitude in $R\rho^{1/d}$, while having nonhyperuniform scaling on larger length scales.
These systems have large and well-defined hyperuniform and nonhyperuniform scaling regimes and thus further exemplify the importance of assessing the degree of order/disorder with respect to a particular length scale \cite{To22, Sk24}.
Our order metrics are able to sensitively detect the changes in order/disorder across length scales for the nearly hyperuniform systems examined herein and we expect them to do so successfully for all other nearly hyperuniform point patterns for which $B/A$ is large (see, e.g., Ref. \citenum{To21} for specific examples).
In addition, at large length scales, we found that the rank ordering of the nonhyperuniform and hyperuniform systems match the rankings of their $B$ and $A$ values, respectively.

We note here that, while the number-variance based scalar order metrics can be applied to any many-particle system in $\mathbb{R}^d$, this is not the case of many other metrics including $s^{(2)}$ and $\tau$, given by Eqs. (\ref{eq:s2}) and (\ref{eq:tau}), respectively.
In particular, $s^{(2)}$ and $\tau$ diverge when applied to systems with long-range order, such as lattices, vacancy-riddled lattices, or quasicrystals.
This weakness can be addressed for the $\tau$ metric by using $\tau(K)$ [given by Eq. (\ref{eq:tauk})]; however, using $\tau(K)$ still requires one to incorporate a range of length scales $[2\pi/K,\infty)$ as opposed to one \textit{specific} length scale.
Moreover, they do not consistently treat antihyperuniform many-particle systems.
For example, the HIP causes these metrics to diverge (i.e., indicating perfect order) because its $g_2(0)$ diverges, while the hard-core antihyperuniform system has $g_2(0)=0$ and is considered by $s^{(2)}$ and $\tau$ to have relatively little order.
Moreover, all of the metrics discussed in Sec. IIC assess the degree of order/disorder of a many-particle system globally (i.e., not at some particular length scale) via some weighted integration of pair statistics, meaning one cannot determine at what specific length scales the system is ordered/disordered. 

Recalling the four properties of a ``good'' order metric from Sec. IIC, it is evident that $\tau$ and $s^{(2)}$ do not satisfy property (1) because of their inconsistent treatment of antihyperuniform systems.
In addition, all existing metrics discussed in Sec. IIC are global assessments of order/disorder and therefore do not satisfy property (3). 
In the present paper, we have shown that the number variance metrics satisfy all four properties of a good order metric set by Kansal {\it et al}. \cite{Ka02} (cf. Sec. II C). Specifically, these metrics can rank the degree of order/disorder of systems with a broad spectrum of short-, intermediate-, and large-$R$ correlations consistently with physical intuition [properties (1) and (2)] and, due to their explicit dependence on a length scale [property (3)], are sensitive to local configurational patterns and the distribution thereof [property (4)].

\textcolor{black}{The features of the $\sigma_N^2(R)/v_1(R)$ curves for our model systems can inform what we should expect from the $\sigma_N^2(R)/v_1(R)$ curves for experimental systems.}
\textcolor{black}{For example, avian photoreceptor patterns \cite{Ji14} have particles with exclusion regions, so one would expect a rapid decay in the small-$R$ regime of $\sigma_N^2(R)/v_1(R)$, like we see in our models with hard-core interactions, and were found to be hyperuniform, so $\sigma_N^2(R)/v_1(R)$ will approach 0 as $R$ approaches $+\infty$.}
\textcolor{black}{In experimental granular packings, such as the shear-jammed disk packings in Ref. \citenum{Zh22}, one would expect a rapid small-$R$ decay in $\sigma_N^2(R)/v_1(R)$ due to the exclusion-volume effects of the disks, as well as oscillations that extend into the intermediate-$R$ regime of $\sigma_N^2(R)/v_1(R)$, due to the presence of coordination shells, such as those observed in the equilibrium hard-particle packings examined here.}
\textcolor{black}{By contrast, in systems that contain aggregates of particles, e.g., protein clusters \cite{Sh16}, one would expect growth in the small-$R$ range of $\sigma_N^2(R)/v_1(R)$, like we observe above in the PCP model.}
\textcolor{black}{In addition, the RVL models are clearly relevant to the well-known occurrence of vacancy defects in crystalline materials, which can be detected via scattering experiments \cite{Pe76,Gi15}, and---like the RVL---will have $\sigma_N^2(R)/v_1(R)$ curves with oscillations at all $R$ and will plateau at some finite positive value proportional to the defect fraction as $R\rightarrow+\infty$.}

To develop more sensitive order metrics, it would be valuable in future work to incorporate higher-order moments of the number of points within a spherical window of radius $R$, such as those formulated in Ref. \citenum{Kl21}.
The number \textit{variance} metrics described in the present paper encode $\rho$ and $g_2(r)$, while higher-order moments as order metrics would encode not only $\rho$ and $g_2(r)$, but also $g_3(r)$, $g_4(r)$, etc., depending on the moment order.
Thus such metrics would additionally be sensitive to the relative distances between triplets, quartets, etc., of particles, allowing for a more comprehensive characterization of order/disorder for particle configurations.
This would allow us to classify the order/disorder of systems with identical pair statistics but different three-body and higher-order correlations including, e.g., the pairs of equilibrium and nonequilibrium configurations with identical pair statistics from Ref. \cite{Wa23} to determine how the degree of order across length scales is affected by how the point pattern is generated.

Another potential area for future research is the application of the metrics devised above to the inverse problem of designing many-particle systems or materials with bulk properties that are tuned via length-specific degrees of order/disorder.
In particular, one could devise a realizable $\sigma_N^2(R)/v_1(R)$ curve with specified small-, intermediate-, and large-scale order/disorder and generate the corresponding point pattern using the techniques described in, e.g., Refs. \citenum{Re07, To09, Di13, Tr20} to produce many-particle systems with desired physical properties including self diffusion coefficients \cite{Br02, Kr09, Kr09_02}, shear and bulk viscosities \cite{Ki49}, isothermal compressibilities \cite{Ha86, To02}, and excess entropies \cite{Ni21}.
One could also consider decorating the points in the pattern with identical nonoverlapping spheres, whose pair statistics are trivially related to the underlying point pattern \cite{To85, To02, To16}, to design packings or granular materials with desired diffusion spreadabilities \cite{To21_02}, dynamic dielectric \cite{To21_03, Ki23, Vy23, Ki24} and elastic constants \cite{Ki20}, fluid permeabilities \cite{To20}, and trapping constants \cite{To20}.

\textcolor{black}{In this work we have focused on the single-argument $\Sigma_N(R)$, which offers an integrated measure of the order/disorder of a many-particle system up to a prescribed length scale $R$.}
\textcolor{black}{The single-argument $\Sigma_N(R)$ is equivalent to the particular choice of $R_i=0$ and $R_j=R$ for the two-argument $\Sigma_N(R_i,R_j)$ and provides a useful way to quantify the overall degree of order/disorder in a many-body system as opposed to $\sigma_N^2(R)/v_1(R)$, which quantifies the order/disorder of a many-particle system at a specific, single length scale.}
\textcolor{black}{Of course, the integral measure $\Sigma_N(R_i,R_j)$ enables one to choose more general arguments in which $0<R_i<R_j$, which would enable one to study integrated measures of the order/disorder of a many-particle system over the specified length scales. Such applications represent fruitful directions for future research.}

\begin{acknowledgments}
The authors thank H. Wang, M. Skolnick, and J. Kim for insightful discussions and valuable feedback on the manuscript. This research was sponsored by the Army Research Office and was accomplished under Cooperative Agreement No. W911NF22-2-0103 as well as the National Science Foundation under Award No. CBET-2133179.
\end{acknowledgments}

\appendix
\section{Direct Sampling of the Number Variance}\label{sec:direct}
For systems that do not have analytical expressions for their pair statistics, we must directly sample the number density fluctuations in a point pattern to determine $\sigma_N^2(R)$, depicted schematically in Fig. \ref{fig:schematic}.
In this work, we follow the two-step procedure used in Ref. \citenum{To21}.
First, observation windows are randomly (i.e., Poission) distributed throughout the point pattern.
Then, we compute the number of points within each observation window using periodic boundary conditions.
To reduce the computational load, the same set of window centers are used for all radii considered in a given point pattern.
The number of windows is chosen such that the volume fraction of their union doed not exceed half the volume of the point pattern $V$, i.e., $1-\textrm{exp}[-N_{window}v_1(R_{max})/V]<0.5$, where $R_{max}$ is the largest window radius considered.
This criterion follows from the exact formula for the volume fraction of Poisson-distributed overlapping spheres \cite{Ch13}.



\begin{thebibliography}{104}
\bibitem [{F1()}]{F1}%
  \BibitemOpen
  \href@noop {} {}\bibinfo {note} {In Ref. \citenum{To22}, the local volume fraction variance $\sigma_V^2(R)$ is used as an order metric across legnth scales for two-phase media and in Ref. \citenum{Sk24}, the local field intensity variance $\sigma_F^2(R)$ is used as an order metric across legnth scales for random scalar fields.}\BibitemShut {Stop}%
\end{thebibliography}
\providecommand{\noopsort}[1]{}\providecommand{\singleletter}[1]{#1}%

\end{document}